\documentclass[12pt,a4paper]{article}
\usepackage[utf8]{inputenc}
\usepackage[T1]{fontenc}
\usepackage[english]{babel}
\usepackage{lmodern}
\usepackage{graphicx}
\usepackage{booktabs}
\usepackage{makecell}
\usepackage{amsmath}
\usepackage{amssymb}
\usepackage{array}
\usepackage{hyperref}
\usepackage{caption}
\usepackage{subcaption}
\usepackage{longtable}
\usepackage{geometry}
\usepackage{enumitem}

\title{A New Method for High-Resolution Dating of Radiocarbon Data: The Example of the First Three Centuries B.C.}
\author{Sebastian~Fürst\\
Universität des Saarlandes\\
Vor- und Frühgeschichte\\
Campus Geb. B3.1\\
66123 Saarbrücken\\
\texttt{sebastian.fuerst@uni-saarland.de}}
\date{}

\begin{document}
\maketitle

\begin{abstract}
Radiocarbon dating poses a challenge in many archaeological contexts due to the limited precision of conventional calibration methods. In this study, we introduce a novel approach to fine-dating that is based on the repeated application of OxCal’s \texttt{R\_Simulate} function. By constructing extensive reference tables and aggregating measures of central tendency (means and medians), uncalibrated 14C measurements are directly mapped to calendar dates. The method is validated through comprehensive simulations and comparisons with dendrochronologically dated tree rings. Despite challenges in segments of the calibration curve with low gradients, the approach demonstrates that a significant improvement in dating precision is achievable. Limitations and potential avenues for further methodological optimisation are discussed.
\end{abstract}


\newpage

\section{Introduction}

Radiocarbon dating is widely recognised as the cornerstone of absolute dating in the archaeological sciences (Libby, 1955; Berger \& Suess, 1979; Aitken 1990; Renfrew \& Bahn 2016). Its transformative impact was underscored by the Nobel Prize awarded to Willard F. Libby in 1960 (Renfrew, 1973; Taylor \& Bar-Yosef, 2014). In recent years, significant advances--–particularly in sample preparation, contaminant removal, and measurement techniques (e.g., Wood, 2015)--–have further refined this method.

Nonetheless, radiocarbon dating remains less effective for certain periods. For instance, intervals characterised by plateaux in the calibration curve--–such as the Hallstatt plateau (approximately 800~BC to 400~BC)--–yield imprecise results (Hamilton et al., 2015; Fahrni et al., 2020; Reimer et al., 2020). Moreover, during periods with robust chronological frameworks established by alternative methods--–for example, the late Iron Age in Central Europe, where typological classifications achieve resolutions of 25 to 50 years (e.g. Barral \& Fichtl, 2012)--–radiocarbon dating has hitherto provided limited additional insights.

Radiocarbon dating’s imprecision stems from two factors: First, the method’s intrinsic uncertainty is reflected in the representation of uncalibrated data as a normal distribution with a given standard deviation (SD). Second, measured radiocarbon ages must be converted into calibrated calendar years (cal~BC or cal~AD) via a calibration curve, whose fluctuating atmospheric 14C levels (see, e.g., Libby, 1955; Libby, 1969; Aitken, 1990) can yield very wide date ranges. Previous attempts to derive precise point estimates from calibrated data have proven insufficient (Michczyński, 2007).

While analysing radiocarbon data from the Roman late Republican military camp at Hermeskeil (Trier-Saarburg, Rhineland-Palatinate) (Fürst \& Hornung, 2024), I encountered the \texttt{R\_Simulate} function in the calibration software OxCal v4.4 (Bronk Ramsey, 1995; Bronk Ramsey, 1998; Bronk Ramsey, 2001; Bronk Ramsey, 2009a). Traditionally employed to validate datasets and optimise sample sizes (Bronk Ramsey, 1995), this function also enables the assessment of dating uncertainties by simulating radiocarbon measurements--–as demonstrated by Bronk Ramsey (2009a) in a Bayesian framework.

Building on these simulations, the question arises: can the \texttt{R\_Simulate} function be harnessed to elucidate the complex relationship between an underlying calendar date and the dispersion of uncalibrated radiocarbon ages from multiple measurements? In practice, repeated measurements of the same object yield a dispersion in radiocarbon ages. By constructing a reference collection of simulated measurements--–each tied to a known calendar date--–we can capture this dispersion and map radiocarbon ages to their corresponding baseline calendar dates.

This reference collection is heuristically valuable, as it establishes a comprehensive mapping between radiocarbon ages and their corresponding baseline calendar dates, thereby enabling the development of a refined estimation procedure for dating archaeological objects. New measurements can be systematically compared with this reference data to retrieve the underlying calendar dates. Because a single radiocarbon age may correspond to multiple calendar dates—resulting in a dispersion around an expected value—it is essential to account for the inherent variances of radiocarbon dating by incorporating at least two, ideally three, independent measurements of the same object (further details on the optimal number of samples are provided in the Results section). Consequently, for an archaeological sample measured multiple times, calculating the mean or median of these normally distributed values yields a robust overall date.

In this paper, we present a novel approach that employs simulated data from OxCal’s \texttt{R\_Simulate} function to construct a reference collection for fine-dating. We detail the construction of this reference collection and outline a statistical estimation procedure--–both in general and in formal mathematical terms--–for approximating target calendar dates. The methodology is validated using dendrochronologically dated tree rings, followed by a comprehensive analysis based on a statistically significant set of test measurements. Finally, we discuss the strengths, limitations, and potential extensions of this approach.

\section{Materials}

\subsection{Software and Computational Methods}

Both the reference collection and the test datasets used to evaluate the method were generated using the online version of OxCal v4.4 (Bronk Ramsey, 1995; Bronk Ramsey, 1998; Bronk Ramsey, 2001; Bronk Ramsey, 2009a). OxCal calibrates uncalibrated 14C ages (reported in BP, with the reference year fixed at 1950) into calibrated calendar dates (expressed in cal~BC/cal~AD) while accounting for uncertainties in both the radiocarbon data and the calibration curve. For the automated extraction of matching radiocarbon ages and subsequent statistical computations, a script was developed in Python 3.10 (Python Software Foundation, 2024), utilising \texttt{pandas} 2.2.2 (McKinney, 2010), \texttt{numpy} 1.24.4 (Harris et al., 2020) and \texttt{openpyxl} 3.1.5 (Gazoni \& Clark, 2023).

\subsection{Functionality of \texttt{R\_Simulate} in OxCal}

For the method presented here, understanding the operation of the \texttt{R\_Simulate} command is essential. In brief, when invoked, \texttt{R\_Simulate} generates a normally distributed random number based on a specified calendar date and a selectable SD. By calibrating this number using the chosen calibration curve, a date is generated that one would expect for an object of a given age, taking into account the achievable accuracy (Bronk Ramsey 2001, 360).

\medskip

\noindent\textbf{Example:} \texttt{R\_Simulate(-200,20)} produces an uncalibrated radiocarbon age with an associated calibrated date range for a calendar date of 200\,BC and a measurement uncertainty of $\pm$20 years.

\medskip

\noindent\textbf{Note:} Even with a nominal SD of zero, method‐inherent fluctuations result in a natural dispersion of radiocarbon ages (see Figures~1a and 1b). Bronk Ramsey (2009a) emphasises that “the measurement is made on the isotopic composition of the sample, not on the age of the sample,” which underlies this dispersion.

\bigskip

\begin{figure}[htbp]
  \centering
  \begin{subfigure}[b]{0.48\textwidth}
    \centering
      \includegraphics[width=\linewidth]{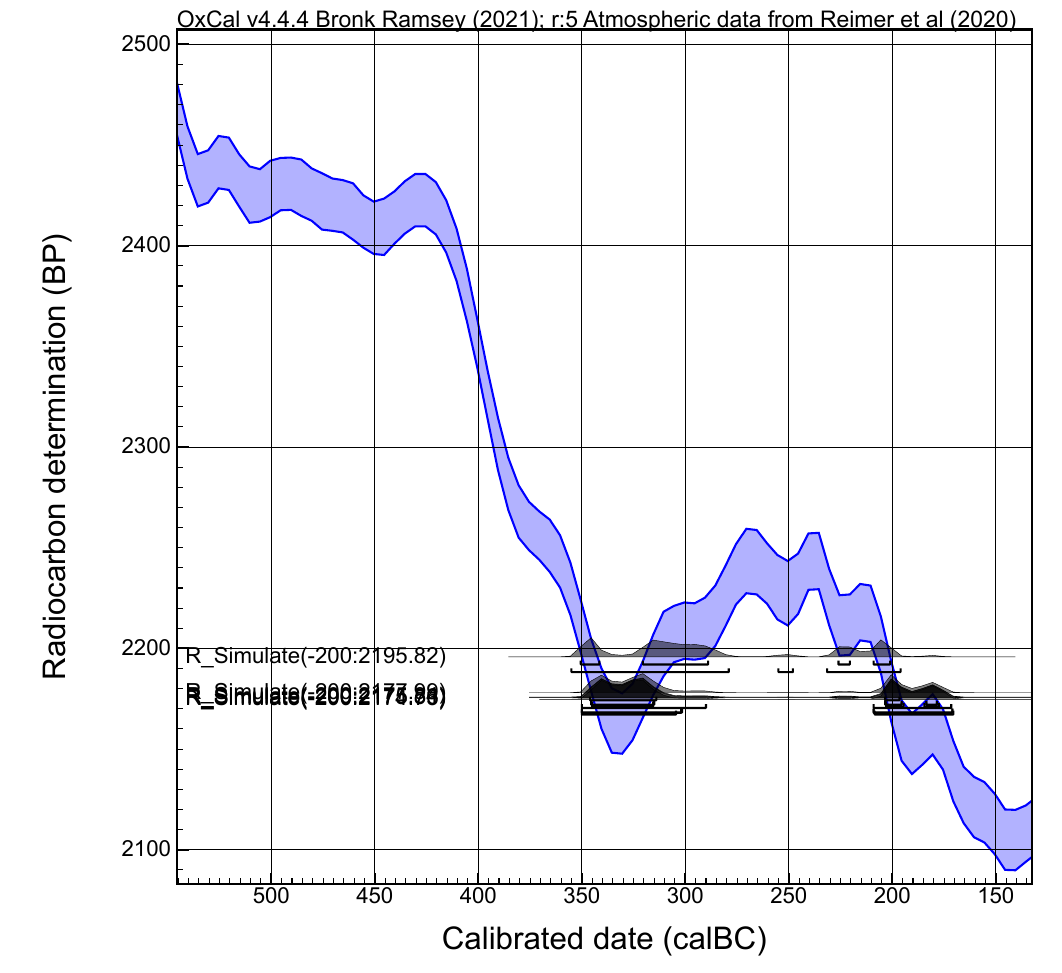}
  \end{subfigure}
  \hfill
  \begin{subfigure}[b]{0.48\textwidth}
    \centering
      \includegraphics[width=\linewidth]{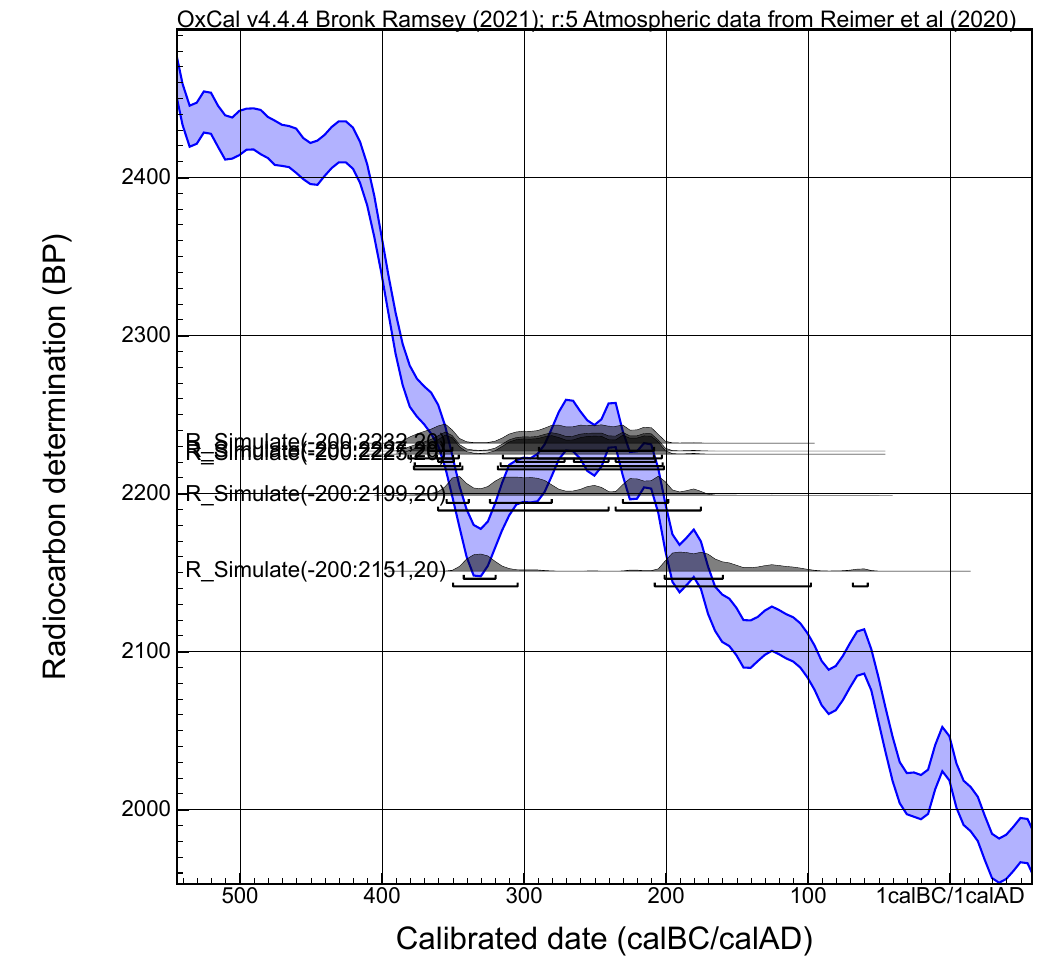}
  \end{subfigure}
  \caption{Distribution of values generated by the \texttt{R\_Simulate} command along the IntCal20 calibration curve--–(a) without SD ($\pm$0 years); (b) with a deviation of $\pm$20 years.}
  \label{fig:fig1}
\end{figure}

\subsection{Structure of the Reference Table}

The reference table was constructed from a series of \texttt{R\_Simulate} computations covering the period from 300\,BC to AD\,20. This time span was selected to capture various forms of the calibration curve within a manageable chronological framework.

By repeatedly simulating radiocarbon ages for fixed calendar dates, the natural, method‐inherent dispersion is replicated. A uniform grid of measurement points--–predominantly at 5‐year intervals (yielding 65 time slices)--–was employed. In addition, variants were generated by varying the number of measurements per time interval and the applied SD (Table~1; each table is included in Appendix~2).

\bigskip

\begin{table}[htbp]
\centering
\caption{Overview of the different reference tables and their respective parameters.}
\label{tab:overview}
\begin{tabular}{l c c c c c}
\toprule
Name & Year interval & Meas./T.Step & SD & Tot. Meas. & Time span \\
\midrule
1\_50\_5   & 1 & 50 & 5  & 10,000  & 1\,BC -- 200\,BC \\
5\_10\_20  & 5 & 10 & 20 & 650     & AD\,20 -- 300\,BC \\
5\_20\_5   & 5 & 20 & 5  & 1,300   & AD\,20 -- 300\,BC \\
5\_50\_5   & 5 & 50 & 5  & 3,250   & AD\,20 -- 300\,BC \\
5\_50\_20  & 5 & 50 & 20 & 3,250   & AD\,20 -- 300\,BC \\
5\_80\_5   & 5 & 80 & 5  & 5,200   & AD\,20 -- 300\,BC \\
5\_100\_0  & 5 & 100 & 0  & 6,500   & AD\,20 -- 300\,BC \\
5\_100\_5  & 5 & 100 & 5  & 6,500   & AD\,20 -- 300\,BC \\
Combo      & 5 & \makecell{20, 50\\80, 100} & 0, 5, 20 & 26,000  & AD\,20 -- 300\,BC \\
\bottomrule
\end{tabular}
\end{table}

\medskip

Several reference table variants were evaluated to assess the impact of temporal resolution and data volume on dating outcomes. For example, the variant 5\_10\_20 (10 measurements every 5 years with a SD of $\pm$20 years) was used initially but later excluded due to inferior performance compared to other variants (e.g. 5\_20\_5 and 5\_50\_5).

To construct the final tables, key outputs from OxCal--–including the precise \texttt{R\_Simulate} command, the individual segments of the 1$\sigma$ and 2$\sigma$ intervals with their probabilities, and the summary statistics that are sigma values, arithmetic means, and medians--–were exported as CSV data. These were subsequently imported into Excel for further processing, where separate columns were created for the underlying calendar dates and the generated 14C ages, and each simulation was assigned an ID.

Figure~2 illustrates a scatter plot of all measurements from the reference table 5\_50\_5, revealing its resemblance to the IntCal calibration curve. Vertically, the simulated 14C ages for each calendar date are approximately normally distributed, as expected from the \texttt{R\_Simulate} computations. Horizontally, the alignment of data points reflects the local gradient and structure of the calibration curve: each horizontal sequence represents repeated simulations at a fixed calendar date, and variations along the x-axis indicate the degree of calibration uncertainty. In regions where the calibration curve flattens--–for example, around 2100 BP--–the horizontal spread is markedly broader, spanning roughly from 160\,BC to 50\,BC, which underscores the potential for multiple calendar date assignments for a single radiocarbon age.

\begin{figure}[htbp]
  \centering
  \includegraphics[width=0.8\textwidth]{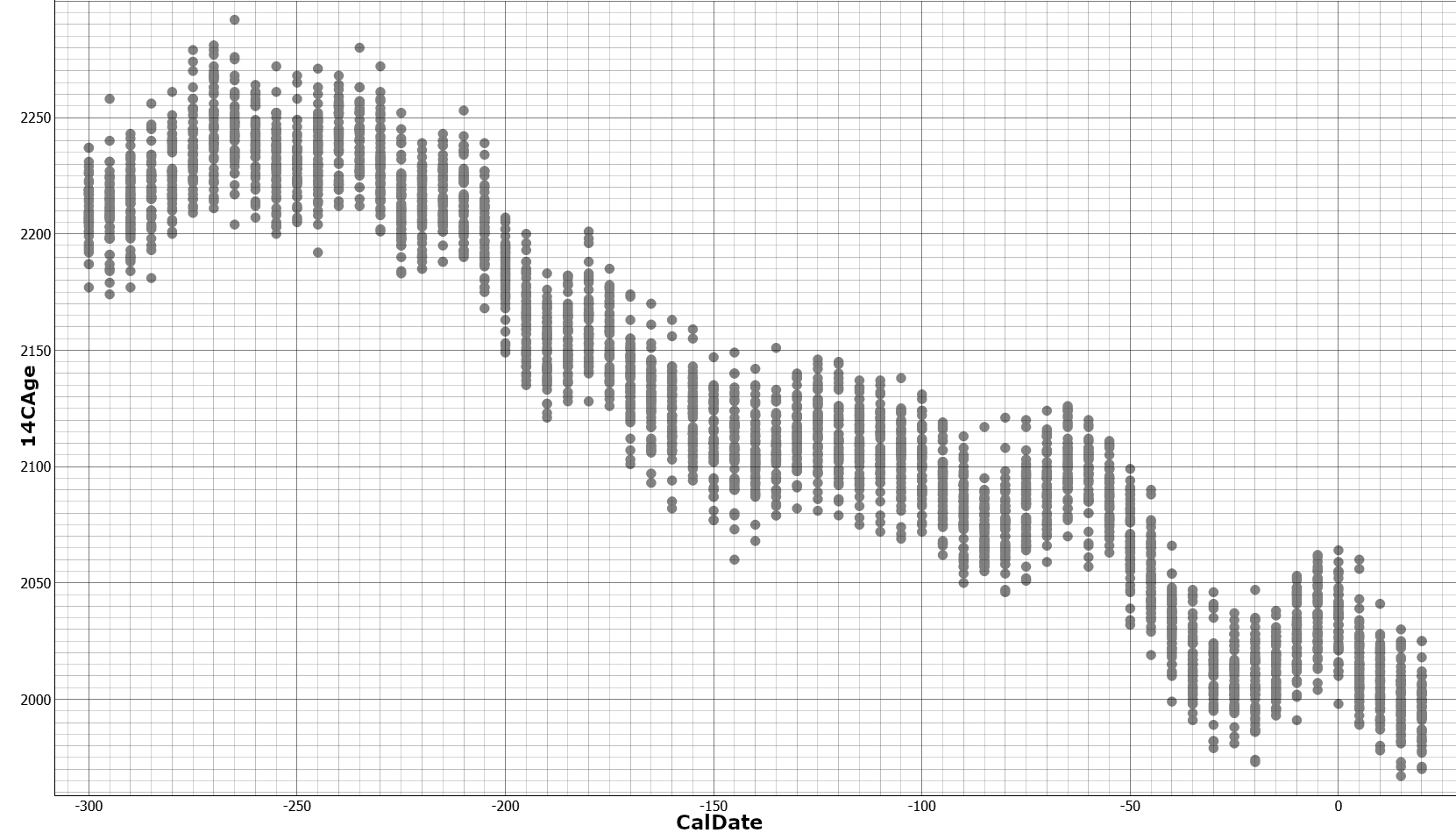}
  \caption{Scatter plot of all measurements from the reference table 5\_50\_5.}
  \label{fig:fig2}
\end{figure}

\section{Methods}

\subsection{Reproducibility and Computational Tools}
To ensure reproducibility, the reference tables and the Python code used for data processing are provided in Appendices 2 and 3, respectively. The reference collection and test datasets were generated using the online version of OxCal v4.4 (Bronk Ramsey, 1995; 1998; 2001; 2009a), which calibrates uncalibrated 14C ages (in BP, reference year 1950) into calendar dates (cal~BC/cal~AD) while accounting for inherent uncertainties. For the automated matching of radiocarbon ages and subsequent statistical analyses, a Python 3.10 script was developed using pandas 2.2.2 (McKinney, 2010), numpy 1.24.4 (Harris et al., 2020), and openpyxl 3.1.5 (Gazoni \& Clark, 2023).

\subsection{Overview of the Dating Method}
This method refines radiocarbon dating by linking multiple measured radiocarbon ages, denoted as $y_i$, of an object to the target calendar date $x$ via simulated calendar dates. A reference table---constructed from numerous OxCal simulations---establishes a multitude of (simulated) relationships between a calendar date and its corresponding radiocarbon age. By filtering the simulated ages $\{y'_{ij}\}$ to retain only those that match the empirical measurements $y_i$, the effective sample size is increased. Moreover, the reference table reveals the underlying simulated calendar dates $\{x'_{ij}\}$, which serve as the basis for computing central tendency measures (e.g., median and arithmetic mean) to estimate the target date $x$. The same tendency measures can also be determined for the means and medians produced in each OxCal simulation and included in the reference table.

\subsection{Variable Definitions and Relationships}

\subsubsection*{Unknown Calendar Date ($x$)}
$x$ is the target calendar date of the object and is the value to be estimated.

\subsubsection*{Number of Measurements}
$n$: the number of radiocarbon measurements performed on the object.

\subsubsection*{Measured Radiocarbon Ages ($y_i$)}
$y_i$ are the uncalibrated radiocarbon ages.

\subsubsection*{Initial multiset:}
\[
Y = \{ y_i \mid i = 1, \dots, m \}
\]

\subsubsection*{Unique values:}
\[
Y^{\text{unique}} = Unique(Y) = \{ y_k^{unique} \mid k = 1, \dots, M \}
\]
where $M$ is the number of unique measurements.

\subsubsection*{Simulated Calendar Dates ($x'_i$) and Radiocarbon Ages ($y'_i$)}
For each simulation, a calendar date $x'_{ij}$ is used to generate a corresponding radiocarbon age $y'_{ij}$ via the \texttt{R\_Simulate} command. The method aims to identify those $y'_{ij}$ that satisfy:
\[
y_i = y'_{ij}.
\]

\subsubsection*{Index Definitions:}
$i$: index of the measured ages $y_i$ ($i = 1, \dots, m$).\\
$j$: index of the simulated ages $y'_{ij}$ matching $y_i$ ($j = 1, \dots, n_i$), where $n_i$ is the number of matches.

\subsubsection*{Relationships between calendar dates and radiocarbon ages:}
\[{x : y = 1 : n}\] 
 A single calendar date \(x\) leads to multiple measured radiocarbon ages \(y_i\) due to measurement uncertainties.
\[{y : y' = 1 : n}\]
Each measured radiocarbon age \(y_i\) corresponds to several matching \(\{y'_{ij}\}\) in the reference table.
\[{y'_{ij} : x'_{ij} = 1 : 1 \text{ (per dataset)}}\]
Within each dataset, each $y'_{ij}$ is linked to a corresponding $x'_{ij}$.
\[{y' : x' = n : m} \text{ (overall relationship)}\]
 Identical $y'$ can result from different $x'$, and conversely, identical $x'$ can lead to different $y'$.\\

\subsubsection*{Explanation of the $n\!:\!m$ Relationship between $y'$ and $x'$ (Table 2):}
In general, there exists an $n\!:\!m$ relationship between the simulated radiocarbon ages $y'$ and the underlying calendar dates $x'$. On the one hand, each $y'$ can arise from different $x'$ because various calendar dates---owing to the method-inherent dispersion---can yield the same value for $y'$. This is why the method requires multiple radiocarbon measurements of the same object. On the other hand, each calendar date $x'$ can produce different $y'$ in each execution of the \texttt{R\_Simulate} command. Consequently, the reference tables were constructed with multiple measurements per time slice.

\subsection{Procedure Description}
\subsubsection*{1. Measurement of Radiocarbon Ages:}
$m$ radiocarbon measurements $y_i$ are performed on the object with unknown date $x$. These measurements yield the radiocarbon ages. Owing to measurement uncertainties, the $y_i$ are approximately normally distributed about the true value corresponding to $x$.

\subsubsection*{2. Creation of the Reference Table:}
For each calendar date $x'_{ij}$ on a uniform grid (e.g., 5-year intervals from 300~BC to AD~20), multiple simulated radiocarbon ages 
\[
y'_{ij} = {R_{Simulate}}(x'_{ij})
\]
are generated, thus replicating the method-inherent dispersion.

\subsubsection*{3. Matching Measured and Simulated Ages:}
For each measured $y_i$, the reference table is queried to extract the corresponding set of calendar dates:
\[
X'_i = \{ x'_{ij} \mid y'_{ij} = y_i \}.
\]

\subsubsection*{4. Combination and Statistical Analysis:}
All matched calendar dates are aggregated:
\[
X' = \bigcup_{i=1}^{m} X'_i.
\]
Statistical measures are then computed directly from $X'_i$ (single-stage approach):

\textbf{Mean of Calendar Dates:}
\[
\bar{x} = \frac{1}{N'} \sum_{i=1}^{m} \sum_{j=1}^{n_i} x'_{ij},
\]
where $N' = \sum_{i=1}^{m} n_i$.

\textbf{Median of Calendar Dates:}
\[
\tilde{x} = \mathrm{Median}(X').
\]
Unique variants are calculated by removing duplicates from $X'$ to reduce bias from over-represented values.

\subsubsection*{5. Usage of Calibration Derived Statistical Measures:}
Each simulated dataset also includes the mean ($\mu_i$) and median ($M_i$) of the calibrated probability distributions. These are collected for each $y_i$ to compute overall means and medians, including a unique variant respectively. Weighted means were tested but found no advantage over unweighted means, due to the prevalence of identical values.

\subsubsection*{6. Summary of the Statistical Measures for Approximating $x$}
\begin{table}[h]
\centering
\small
\begin{tabular}{lll}
\toprule
\textbf{Measure} & \textbf{Abbreviation in Tables} & \textbf{Calculation} \\
\midrule
Mean of Calendar Dates (CDs) & CalDate Mean & $\hat{x} = \frac{1}{N'} \sum x'_{ij}$ \\
Median of CDs & CalDate Median & $\tilde{x} = \mathrm{Median}(X')$ \\
Mean of Unique CDs & unique\_CalDate Mean & $\hat{x}^{\text{unique}} = \frac{1}{N_{\text{unique}}} \sum x^{unique}_{j}$ \\
Median of Unique CDs & unique\_CalDate Median & $\tilde{x}^{\text{unique}} = \mathrm{Median}(X'^{unique})$ \\
Mean of Means & Mean Mean & $\hat{\mu} = \frac{1}{N'} \sum \mu_{ij}$ \\
Median of Means & Mean Median & $\tilde{\mu} = \mathrm{Median}(\{\mu_{ij}\})$ \\
Mean of Unique Means & unique\_Mean Mean & $\hat{\mu}^{unique} = \frac{1}{N_{\mu, unique}} \sum \mu_{j}^{unique}$ \\
Median of Unique Means & unique\_Mean Median & $\tilde{\mu}^{unique} = \mathrm{Median}(\{\mu_{j}^{\text{unique}}\})$ \\
Mean of Medians & Median Mean & $\hat{M} = \frac{1}{N'} \sum M_{ij}$ \\
Median of Medians & Median Median & $\tilde{M} = \mathrm{Median}(\{M_{ij}\})$ \\
Mean of Unique Medians & unique\_Median Mean & $\hat{M}^{unique} = \frac{1}{N_{M,unique}} \sum M_j^{unique}$ \\
Median of Unique Medians & unique\_Median Median & $\tilde{M}^{unique} = \mathrm{Median}(\{M_j^{unique}\})$ \\
\bottomrule
\end{tabular}
\caption{Summary of the statistical measures for approximating $x$.}
\end{table}

\subsection{Key Assumptions and Practical Conditions}
\begin{itemize}
    \item The measurements $y_i$ are independent and normally distributed.
    \item The uncertainties of both measurements and simulations are known.
    \item The reference table contains exact matches between the measured $y_i$ and the simulated $y'_{ij}$.
    \item Both the measured $y_i$ and the simulated $y'_{ij}$ are assumed to be normally distributed.
    \item The use of multisets ensures that the multiplicity of identical values is properly accounted for, thereby mitigating bias.
\end{itemize}

\subsection{Practical Implementation of Absolute Dating}
When applying the method to an archaeological object, it is essential to consider the general sources of uncertainty inherent in radiocarbon samples (see, for example, Aitken, 1990, pp. 86--92; Higham \& Petchey, 2000, p. 265, Table 4; Bayliss, 2009, p. 129; Bayliss, 2015, p. 690; Hamilton \& Krus, 2018, pp. 194, 198). Factors such as sample selection (preferably objects with short inherent lifespans) and cost--benefit considerations (notably, that three measurements generally yield robust results) are essential for practical application. Such considerations, however, vary on a case-by-case basis and should be evaluated and adjusted individually (see, for example, Bronk Ramsey, 1995; Holland-Lulewicz \& Ritchison, 2021).

To validate the method, test datasets can be generated using \texttt{R\_Simulate}. Crucially, the original calendar date used in the simulations is not employed to compute the statistical indicators (which is the purpose in the reference tables) but rather serves as a control to assess the degree of deviation in specific time intervals. In order to obtain a realistic simulation of an empirical fine-dating scenario, three individual measurements based on the same underlying calendar date should be used as the test dataset (see the Results chapter). Test series comprising at least 100 test datasets per time interval (i.e. 300 individual measurements) are particularly well-suited, as they provide a statistically significant sample size. For this purpose, the same OxCal-lists of \texttt{R\_Simulate} commands used to generate the reference tables may be employed, since each execution of the code produces new random numbers.

With the aid of a simple Python script (see Appendix 3), the individual \texttt{R\_Simulate} computations exported from OxCal as a CSV file can then be grouped by calendar year into clusters of three calculations each and saved as separate CSV files. In this way, numerous test computations based on three individual measurements with the same base date can be generated quickly and efficiently. Only the designation \texttt{R\_Simulate} and the specification of the underlying calendar date and its associated standard deviation differ from the traditional \texttt{R\_Date} of a classical calibration; however, these modifications can be easily implemented using a Python script or by employing “find and replace” commands in Excel.

When applying the fine-dating method--–whether using a simulated test dataset or “real” empirical measurements from a research context--–the identical radiocarbon ages must be located within a reference table, as mentioned above. Both this search and the subsequent calculation of the 12 statistical indicators can be automated. For this purpose, another Python script has been developed (see Appendix 4). Fundamentally, however, these steps can also be performed manually without recourse to an algorithm, as was done at the outset of this study. In terms of workflow, the procedure is therefore very low-threshold.

Figure 3 presents a schematic overview of the method applied to a single archaeological object with a target calendar date of 75~BC. In this example, the object was dated three times, yielding three distinct 14C ages (2073~BP, 2087~BP, and 2097~BP, each with a standard deviation of ±20 years). These measurements were exported from OxCal as a CSV file, and the Python script subsequently automates the matching of these 14C ages with the entries in the reference table. In this case, the reference table (variant 5\_20\_5, comprising 20 measurements at 5‐year intervals with a standard deviation of ±5 years) produced 21 matching records. The Python script then generates an Excel file containing two worksheets (see Appendix~5):
\begin{itemize}
    \item The first worksheet, titled “Overview”, lists each matched record from the reference table along with the calendar date underlying the \texttt{R\_Simulate} computation and includes the mean, median, and sigma of the corresponding calibrated probability distribution. Additional index columns ensure transparency and traceability of data sources.
    \item The second worksheet, consistently labelled “Summary” throughout the study, contains the 12 statistical indicators (both means and medians, including their unique variants) derived from the matched datasets.
\end{itemize}

\begin{figure}[htbp]
  \centering
  \includegraphics[width=0.8\textwidth]{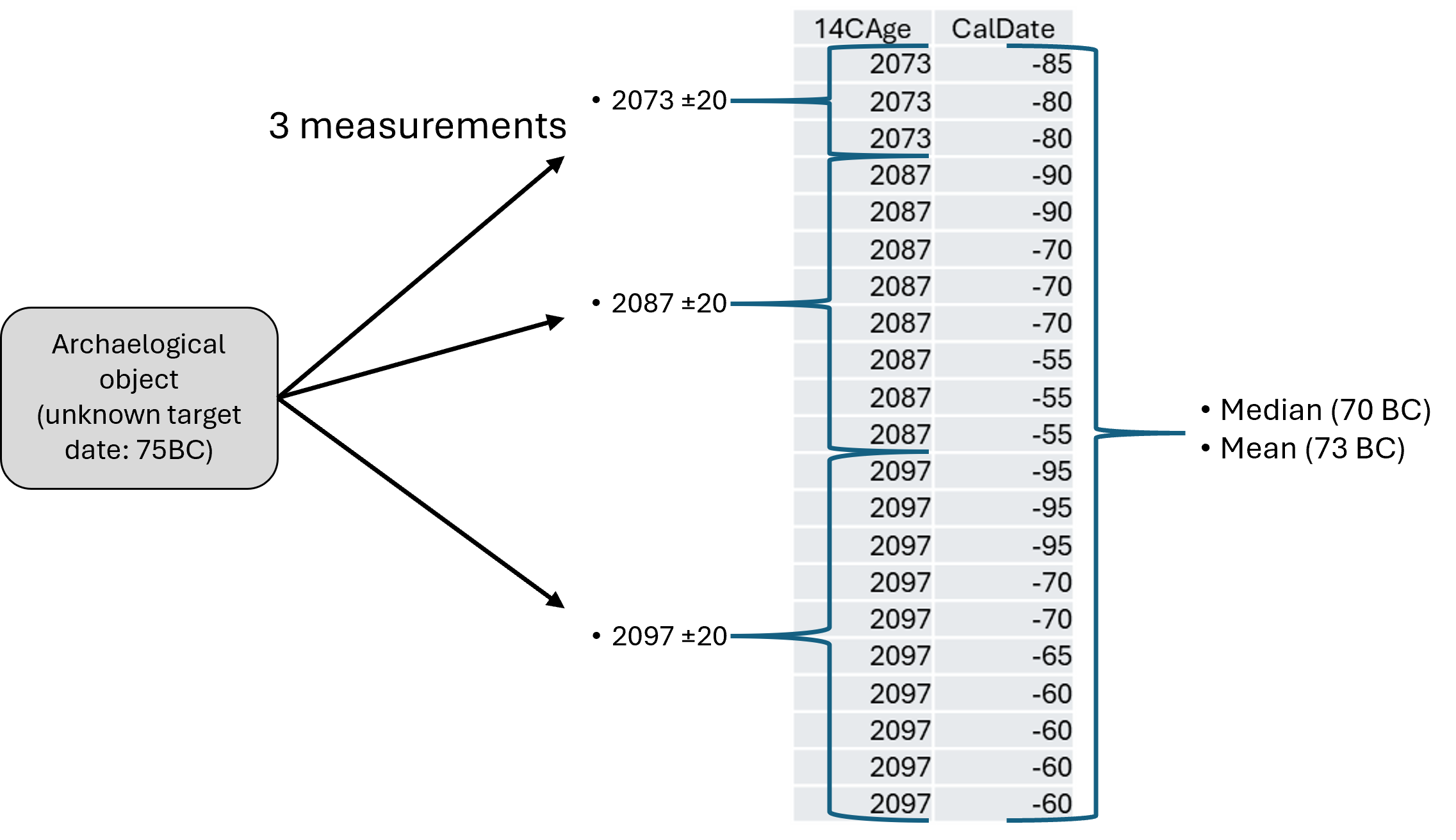}
  \caption{Schematic of the estimation procedure illustrated by three radiocarbon dates from a sample with the target date 75~BC.—Reference table employed: 5\_20\_5 (20 measurements at 5-year intervals, from 300~BC to AD~20, with a standard deviation of ±5 years).}
\end{figure}

This practical implementation demonstrates the method’s strength: by linking a limited set of measured radiocarbon ages with an extensive reference table, a suite of statistical indicators can be computed to reliably estimate the target calendar date. Furthermore, the low-threshold nature of the procedure—--based on readily available software and Python scripts—--facilitates its application not only in archaeological contexts but also in other disciplines that utilise radiocarbon dating.

\subsection{Evaluation of the Dating Results}
In terms of practical application, one further aspect remains outstanding: an evaluation of the results that can indicate whether the statistical parameters computed by the fine-dating method have indeed provided an accurate approximation of the target calendar date---particularly given that the available radiocarbon ages are derived from a non-problematic segment of the calibration curve. Moreover, it would be advantageous to assess which of the 12 parameters yields superior results in cases of larger discrepancies.

A veritable plethora of statistical--mathematical methods exists for this purpose (cf. the Discussion section); however, these have been deliberately disregarded in this paper, as the focus is solely on demonstrating the method for computing the 12 indicators.

Fundamentally, the shape of the calibration curve in the region to be dated is decisive for the quality of the estimates. In this respect, the method does not differ from conventional calibrations. In regions where the calibration curve is relatively flat---the so-called plateaus---this method, too, is expected to yield less precise results (see, for example, Fahrni et al., 2020; Holland-Lulewicz \& Ritchison, 2021).

A straightforward and easily implementable preliminary assessment of the measurement results can be achieved by employing a test series based on the \texttt{R\_Simulate} command. In this approach, a statistically significant number of test datasets is subjected to fine-dating, ensuring that for each dataset the 12 statistical indicators are available. The basic principle is analogous to the fine-dating process itself: a collection of simulated measurements is generated to serve as a reference against which empirically obtained data can be compared. In this instance, however, both the simulated and the empirical data have already undergone the aforementioned fine-dating process, so that for each dataset---whether in the reference table or among the empirical data---all 12 statistical indicators are available.

Subsequently, for each reported statistical value $x_i$ (for example, “CalDate Median”) derived from an empirical fine-dating (see Appendix~6a, hereafter referred to as the report file, as the data originate from the fine-dating report), we first extract the corresponding set of values $R$ from the reference table (Appendix~6b). A Python script has been developed for this purpose (see Appendix~6c).

In the script, the tolerance range is initially set to ±1 and is incrementally increased—--if fewer than five matches are found—--up to a maximum of ±10. From the matched datasets, the calendar date that occurs most frequently (the “Most Probable Date”) and the range of these values are computed. The latter enables an assessment of whether a wide dispersion exists, which may indicate problematic regions of the calibration curve. If the column “OriginalCalDate” is present in the report file (in the case of simulated data), differences (delta values) between the calculated “Most Probable Date” and the original value are additionally determined. Such differences can be used to gain an overview of the calibration curve’s behaviour across different time intervals. Finally, all results are saved in a new Excel file.

The method is based on a tolerance‐based search, whereby for each $x_i$ we define the set of reference values that lie within a specified tolerance:

\[
M_i(T) = \{ r \in \mathbb{R} \mid |x_i - r| \le T \}
\]
where $T$ is set to a starting value $T_0$ (e.g., $T_0 = 1\,\text{year}$). If the number of matches is insufficient, i.e.,
\[
\left| M_i(T) \right| < m_{\min},
\]
then $T$ is incremented by a fixed step $\Delta T$ until either
\[
\left| M_i(T) \right| \geq m_{\min}
\]
or the maximum tolerance $T_{\max}$ is reached.

Once an appropriate tolerance $T_i$ is determined for $x_i$, two key quantities are computed:

\textbf{1. Most Probable Date (MPD)}\\[1ex]
This is defined as the mode of the matching set:
\[
\text{MPD}(x_i) = \operatorname{mode}\left(M_i(T_i)\right).
\]
\textbf{2. Overall Range:}\\[1ex]
Defined as the difference between the maximum and minimum values in $M_i(T_i)$:
\[
R(x_i) = \max\{ r \mid r \in M_i(T_i) \} - \min\{ r \mid r \in M_i(T_i) \}.
\]
If an original calendar date $O_i$ (from the report file) is available for $x_i$, a delta value can be computed as:
\[
\delta(x_i) = \text{MPD}(x_i) - O_i.
\]

\subsubsection*{Aggregating the Results:}
Finally, to provide a single aggregated estimate of the target calendar date, we calculate summary statistics over the entire dataset. Specifically, the Overall Mean is computed as the average of all MPD values:
\[
\text{Overall Mean} = \frac{1}{N} \sum_{i=1}^{N} \text{MPD}(x_i),
\]
and the Overall Median is defined as the median of the MPD values:
\[
\text{Overall Median} = \operatorname{median}\{ \text{MPD}(x_1), \text{MPD}(x_2), \ldots, \text{MPD}(x_N) \}.
\]
These overall metrics serve to summarise the fine-dating results across the entire dataset, offering a robust aggregate estimate that accounts for variability among the individual statistical indicators.

\subsection{Lookup table}
These additional statistical indicators, however, yield values only for individual measurements\allowbreak---whether derived from “real” radiocarbon analyses of archaeological objects or from simulated data that enhance our understanding of performance across various segments of the calibration curve. To asses the quality of a given indicator, a dedicated Python script was developed (see Appendix 7). When executed on simulated data, this script constructs a lookup table that quantifies the percentage of reference measurements falling within a specified standard deviation. In the present study, tolerance intervals of ±12 years (approximately one generation, commonly encountered in fine chronological phases) and ±25 years were selected. Consequently, for each absolute dating, the lookup table can be consulted to determine which statistical indicator is most likely to yield a precise date.

To achieve this, the script defines class intervals in 5‐year increments for each individual statistical indicator (e.g. CalDate Median or CalDate Mean). Each indicator is processed independently, so that, for instance, a measurement with a CalDate Median of –251 is assigned to the interval
\[I_j=[-255,\ -250),\]
whereas the same measurement, if considered under CalDate Mean, might be allocated to a different interval, such as
\[I_{j\prime}=[-245,\ -240).\]
For an indicator k, we partition the range into equally spaced intervals (in 5‐year steps):

\[I_1=[-300,\ -295),\ I_2=[-295,\ -290),\ \ldots,\ I_m=[-5,\ 0).\]
A reported value $k_i$ is assigned to the interval $I_j$ that contains it. Within each interval $I_j$, the number of occurrences where the deviation $\Delta$ (for example, $\Delta =\hat{x}-x_{orig}$) falls within ±12 (or ±25) is counted. Formally, for indicator k and interval $I_j$, the fraction is given by:

\[
\text{Fraction}_{\pm12}(I_j,k) = \frac{\#\{ k_i \in I_j : \Delta_i \le 12 \}}{\#\{ k_i \in I_j \}} \times 100\%
\]
with a similar expression for ±25.

Using these counts, the script creates a lookup table in which the rows correspond to the class intervals $I_j$ (identified by their left boundary, denoted as “BucketLeft”) and the columns represent the individual indicators $k$. For each indicator $k$ and interval $I_j$, the table reports:
\begin{itemize}
  \item $\text{TotalCount}_k(I_j)$: the total number of records in $I_j$;
  \item $\text{Fraction}_k^{(\pm12)}(I_j)$: the percentage of records with $\left|\mathrm{\Delta}\right|\le 12$;
  \item $\text{Fraction}_k^{(\pm25)}(I_j)$: the percentage of records with $\left|\mathrm{\Delta}\right|\le 25$.
\end{itemize}

Thus, if a particular measurement exhibits a CalDate Median of $\ -251$, one would consult the lookup table for the CalDate Median indicator in the interval $[-255,\ -250)$ to read off the corresponding percentages. Similarly, if the same measurement yields a CalDate Mean of $\ -242$, the relevant percentages would be found in the interval $[-245, \ -240)$ of the CalDate Mean column. In this way, the lookup table provides a rapid and systematic means of mapping each computed indicator to its corresponding percentage deviation $\Delta$ within defined ranges—even when different indicators fall into distinct intervals.

\section{Results}

The high-resolution dating method enables a precise estimation of an object's unknown calendar date by statistically analysing multiple radiocarbon measurements and comparing them with a reference table. The combination of means, medians, and unique variants minimises bias and maximises the accuracy of the estimate. Moreover, the method accounts for the multiplicity of data points, thereby reducing potential distortions arising from repeated values.

\subsection{Method Validation}

\subsubsection{Test Series 1: Simulation of an Object with a Calendar Date of 75 BC}

Since the computations presented in the Methods section constitute a full simulation---where the test data were generated via a random number generator---the results can be discussed in aggregate. To illustrate the performance of the fine-dating method, we evaluated the deviations (deltas) of the twelve primary statistical indicators computed by the method for three simulated 14C ages (2073~BP, 2087~BP, and 2097~BP). This example clearly demonstrates the importance of performing at least three measurements; such a practice effectively compensates for statistical outliers (e.g. the 14C age of 2097~BP) while keeping costs in check. Although additional measurements might capture the natural dispersion more accurately, our test calculations revealed only marginal improvements beyond three measurements.

To present these results succinctly, we have aggregated the individual delta values into a summary table (Table 2, Aggregated Evaluation). This table reports, for each indicator type, the average delta and the range (i.e. the minimum and maximum deviations) relative to the target base date (75~BC).

\medskip

\begin{table}[]
\centering
\begin{tabular}{l r r}
\toprule
\textbf{Indicator} & \textbf{Avg. $\Delta$ (yrs)} & \textbf{$\Delta$ Range (yrs)} \\
\midrule
CalDate Median & 5 & 5 -- 8 \\
CalDate Mean   & 2 & 0 -- 7 \\
Unique CalDate Median & 0 & 0 -- 8 \\
Unique CalDate Mean   & 0 & -1 -- 8 \\
Mean Median & -17 & -29 -- 3 \\
Mean Mean   & -21 & -29 -- 3 \\
Unique Mean Median & -23 & -29 -- 3 \\
Unique Mean Mean   & -20 & -30 -- 3 \\
Median Median & -14 & -29 -- 5 \\
Median Mean   & -20 & -29 -- 5 \\
Unique Median Median & -21 & -29 -- 5 \\
Unique Median Mean   & -20 & -29 -- 5 \\
    \bottomrule
    \end{tabular}
    \caption{Aggregated summary of the average deviations of the twelve statistical indicators for the three 14C ages relative to the target date (75\,BC). The reference table employed (variant 5\_20\_5) comprises 20 measurements at 5-year intervals from 300\,BC to AD\,20, with a SD of $\pm$5 years.}
    \label{tab:my_label}
\end{table}

This aggregated summary clearly indicates that the indicators based directly on calendar dates (i.e. CalDate Median and CalDate Mean, along with their unique variants) yield minimal deviations, whereas those computed from the means and medians of the probability distributions exhibit larger negative deviations. Such findings support the conclusion that direct calendar date computations produce more reliable estimates under the conditions tested.

\subsubsection{Test Series 2: Dendrochronologically Dated Samples}

To validate the method, four dendrochronologically dated tree rings from Stuiver et al. (1998) were selected, with the following radiocarbon ages and SDs:

\medskip

\begin{tabular}{l l}
Dendrochronological Date & 14C Age ($\pm$ SD) \\
\midrule
15\,BC & 1999 $\pm$ 10\,BP \\
16\,BC & 2003 $\pm$ 16\,BP \\
25\,BC & 2022 $\pm$ 16\,BP \\
26\,BC & 2035 $\pm$ 15\,BP \\
\end{tabular}
\medskip

The mean of these calendar dates is 20.5\,BC, which in this example represents the (otherwise unknown) target event date \( x \). Although computing a mean from several independently dendrochronologically dated tree rings is acceptable, in typical archaeological contexts it is essential to ensure that the measurements pertain to a single object. Further computations for dendrochronologically dated tree rings are provided in Appendix~8.

\subsubsection*{Execution of the Calculations}

The four available 14C ages, denoted \( y_1 \) to \( y_4 \), were first matched against the reference table 5\_100\_5, which comprises 100 measurements at 5‐year intervals (from 300\,BC to AD\,20) with a SD of $\pm$5 years. This matching process yielded a total of 79 records with 7 matches for 1999\,BP, 22 for 2003\,BP, 29 for 2022\,BP, and 21 for 2035\,BP (see Appendix~8 for the complete list), thus expanding the dataset by a ratio of approximately 1:20. By extracting the repeatedly occurring values from the reference table, a substantially larger basis for computing means and medians is obtained.

The corresponding calendar dates for these 79 records, which range from 50\,BC to AD\,20, allow us to infer the underlying calendar dates for the four tree rings. Notably, these 79 values are unevenly distributed across 15 distinct calendar dates, with two clusters emerging and a gap around 20\,BC—corresponding to the computed mean of the tree-ring calendar dates---while higher values cluster near 30\,BC and 15\,BC. The frequency distributions for calendar dates, medians, and means are summarised in Figure~4. In particular, the medians and means exhibit lower variance; the 79 records are grouped into seven distinct medians (spanning from 16\,BC to AD\,10) and six distinct means (ranging from 19\,BC to AD\,6), with a tendency for the central values to be skewed towards the earlier part of the intervals.

\begin{figure}[htbp]
\centering
  \begin{subfigure}[b]{0.48\textwidth}
    \centering
      \includegraphics[width=\linewidth]{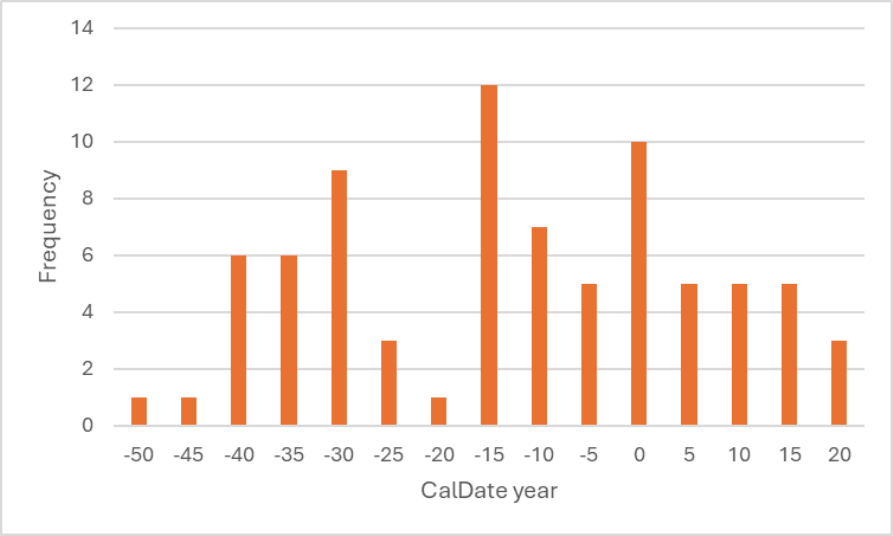}
          \subcaption{Calendar dates}
    \label{fig:4a}
   \end{subfigure}
        \hfill
  \begin{subfigure}[b]{0.48\textwidth}
    \centering
      \includegraphics[width=\linewidth]{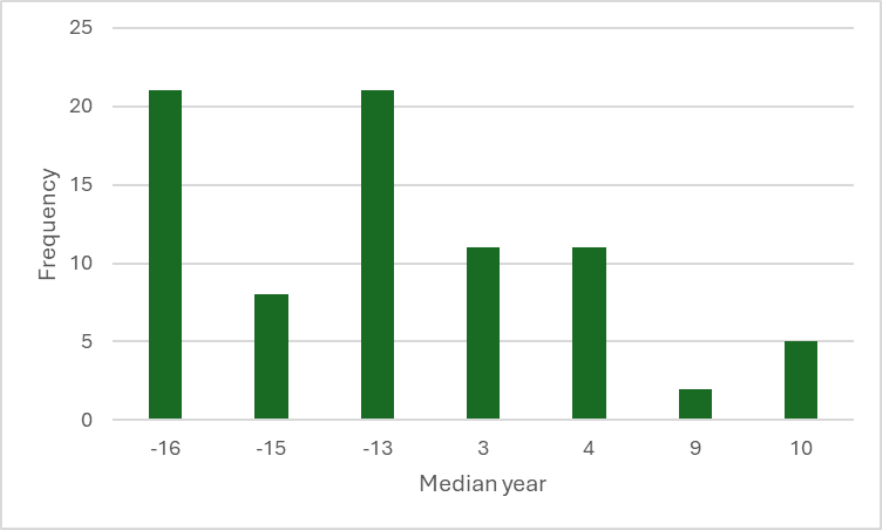}
          \subcaption{Medians}
            \label{fig:4b}
    \end{subfigure}
\vspace{1ex}
 \begin{flushleft}
  \begin{subfigure}[b]{0.48\textwidth}
        \raggedright
      \includegraphics[width=\linewidth]{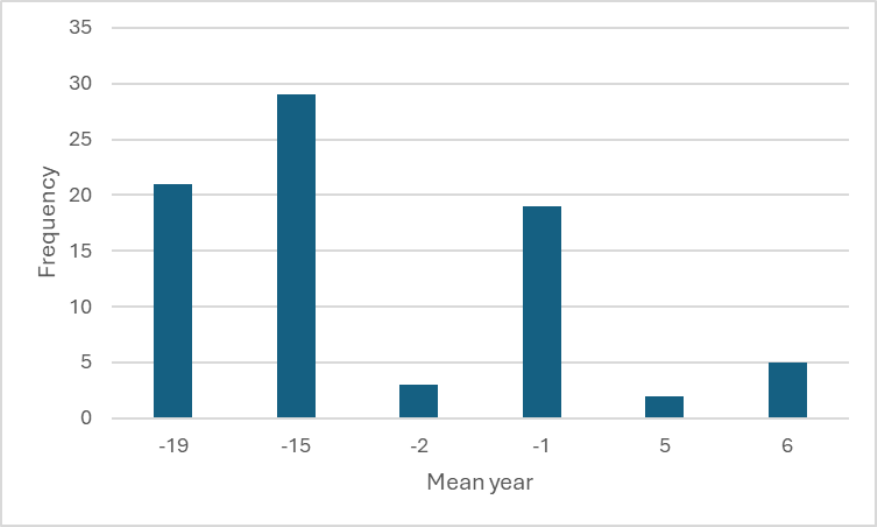}
            \subcaption{Means}
      \label{fig:4c}
  \end{subfigure}
\end{flushleft}
\caption{Frequency distributions of the three primary parameters for the fine-dating method; \( n = 79 \). (a) Calendar dates; (b) Medians; (c) Means.}
\label{fig:fig4}
\end{figure}

Aggregated computations of the means and medians (and their unique variants) produced estimates of the unknown calendar date \( x \) that deviated by an average of only 11.9 years, with no individual value differing by more than 23.5 years (Table~3). Notably, the unique variants of the calendar dates yielded particularly accurate approximations, approaching the target value within as little as 5.5 years. In contrast, a conventional two-stage procedure---where an estimate \( \hat{x_i} \) is computed for each \( y_i \) prior to deriving a combined mean---resulted in larger deviations (approximately 15.5 years for the mean and 11 years for the median).

For clarity, an aggregated summary of these results is presented in Table~3 (detailed values are provided in Appendix~8):

\medskip
\begin{table}[h]
  \centering
\begin{tabular}{l r r}
\textbf{Indicator} & \textbf{Avg.$\Delta$ (yrs)} & \textbf{$\Delta$ (20,5~BC)} \\
\midrule
CalDate Median & -10.00 & 10.5 \\
CalDate Mean   & -12.34 & 8.2 \\
Unique CalDate Median & -15.00 & 5.5 \\
Unique CalDate Mean   & -15.00 & 5.5 \\
Mean Median    & -15.00 & 5.5 \\
Mean Mean      & -10.37 & 10.1 \\
Unique Mean Median & -1.50 & 19.0 \\
Unique Mean Mean   & -4.33 & 16.2 \\
Median Median  & -13.00 & 7.5 \\
Median Mean    & -7.39  & 13.1 \\
Unique Median Median & 3.00  & 23.5 \\
Unique Median Mean   & -2.57 & 17.9 \\
\bottomrule
\end{tabular}
  \caption{Aggregated overview of the average deviations (deltas) of the statistical indicators for the four tree-ring samples using reference table 5\_100\_5.}
\end{table}

\medskip
This aggregated summary clearly demonstrates that the method yields robust estimates of the target calendar date, with the unique variants of the calendar dates providing the most precise approximations.

\subsubsection{Test Series 3: Simulations on a Statistically Significant Scale}

Following successful validation of the fine‐dating method using dendrochronologically dated measurements from real research contexts, this section evaluates the method’s performance on a statistically significant scale, while considering the respective advantages and limitations of the various reference tables.

\subsubsection*{Simulated Test Data}
Simulations were conducted over 61 time intervals (at 5‐year increments) spanning 300\,BC to 0 on the floating point timescale, with 100 test datasets per interval. Each dataset comprises three \texttt{R\_Simulate} computations based on the respective calendar date and a SD of $\pm$20 years, amounting to 18,300 individual computations. This extensive simulation framework permits a robust statistical analysis of the method’s parameters (see Appendix~9 for full details).

\subsubsection*{Statistical Dispersion and Normality}
An initial inspection of the test data is provided by the distribution of 300 \texttt{R\_Simulate} computations corresponding to the time point 0 (see Figure~5 and “IndividualComputations” in Appendix~9). These computations yield 94 distinct 14C ages, indicating a unique-to-total ratio of approximately 1:3 and confirming that the \texttt{R\_Simulate} command’s random number generator produces a normally distributed dispersion of values.

\begin{figure}[htbp]
  \centering
  \includegraphics[width=0.8\textwidth]{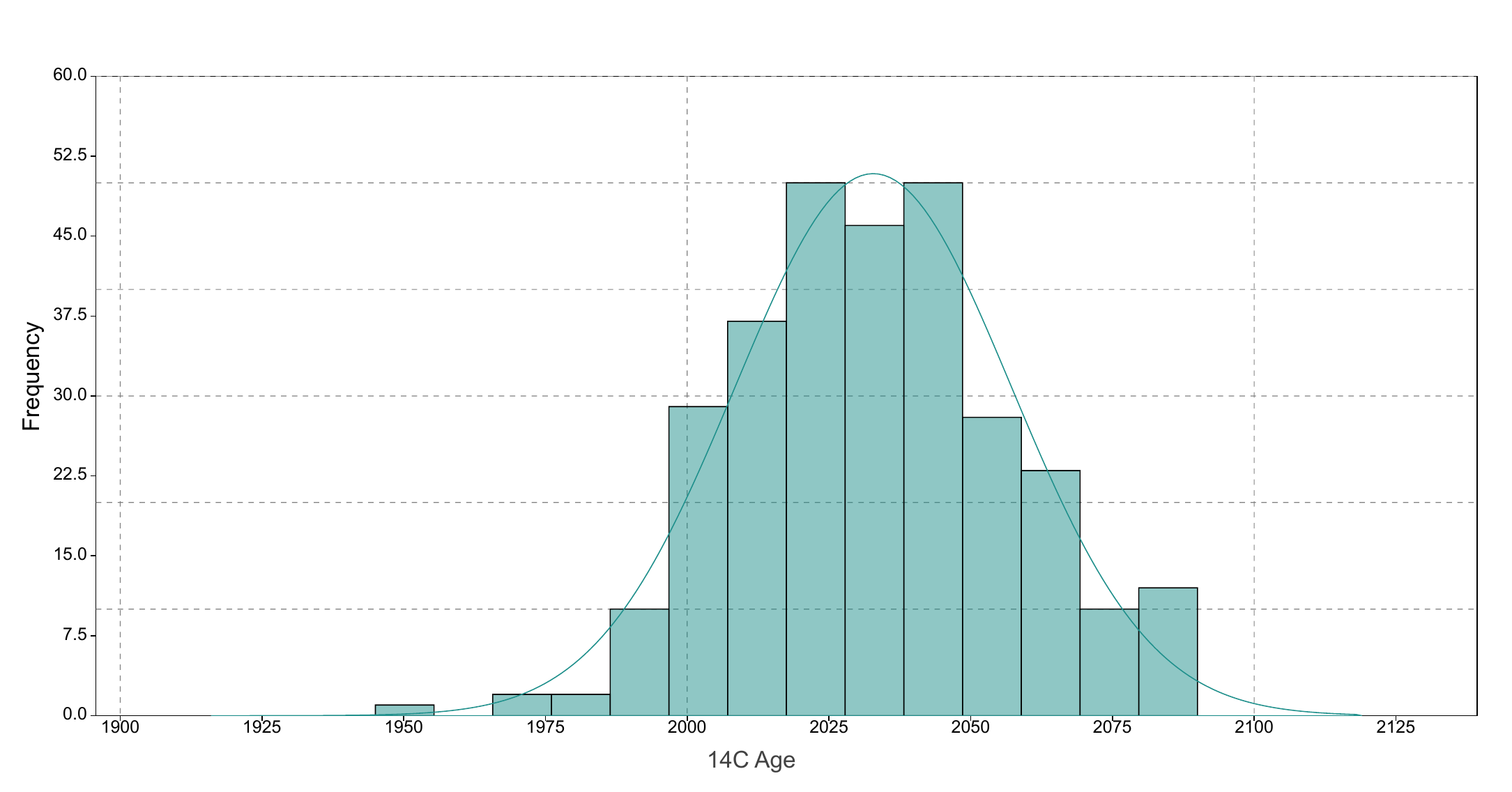}
  \caption{Histogram of the 14C ages generated by \texttt{R\_Simulate} in Test Series 3 for the time point 0. It is clearly evident that the data are normally distributed; \( n = 300 \); 14 bins according to the Rice rule.}
  \label{fig:fig5}
\end{figure}

To rigorously assess normality, the D’Agostino–Pearson test (D’Agostino, 1971; D’\allowbreak Agostino \& Pearson, 1973) was applied to each of the 61 intervals using the \texttt{normaltest} function from the Python SciPy library (Virtanen et al., 2020). The test evaluates the null hypothesis that the 14C ages for a given interval are drawn from a normally distributed population.

In our analysis, the p-value exceeded 0.05 in all but four intervals (20, 25, 195, and 300\,BC); for example, the interval at 0 exhibited a p-value of 0.86 (for full documentation see worksheet 1 in Appendix~10). Although tests with sample sizes \( n > 200 \) may be overly sensitive (Demir, 2022), the overall data indicate only minor departures from normality. In summary, the 6,100 simulated tests exhibit a predominantly normal distribution (mean test statistic: 1.81; average p-value: 0.53).

\subsubsection*{Ratios and Variability Using Reference Table 5\_20\_5:}
Focusing on the time interval corresponding to 0 on the floating point timescale, matching between the test measurements (\( y_{(1-300)} \)) and reference table 5\_20\_5 yielded 1,163 matches (see worksheet~2 in Appendix~9). Considering that each of the 100 independent tests (each comprising three measurements) is treated separately, duplications are inevitable. Analysis reveals that out of 92 unique values derived from the 300 measurements, only 82 unique values are present in the extracted set from the reference table (i.e. $y'^{unique}_{ij} = 82$), with five 14C ages---ranging between 1945 and 1989\,BP---falling outside the range of the applied reference table (for full documentation see worksheet 2 in Appendix~10).

This observation underscores the necessity of incorporating an adequate buffer zone when constructing reference tables; in this instance, the buffer extends only to AD 20, which appears insufficient to encompass more remote statistical outliers. Moreover, the relatively low data volume (20 measurements per interval) in the reference table increases the risk that certain values may not be represented, potentially compromising the results.

\begin{figure}[htbp]
  \centering
  \includegraphics[width=0.8\textwidth]{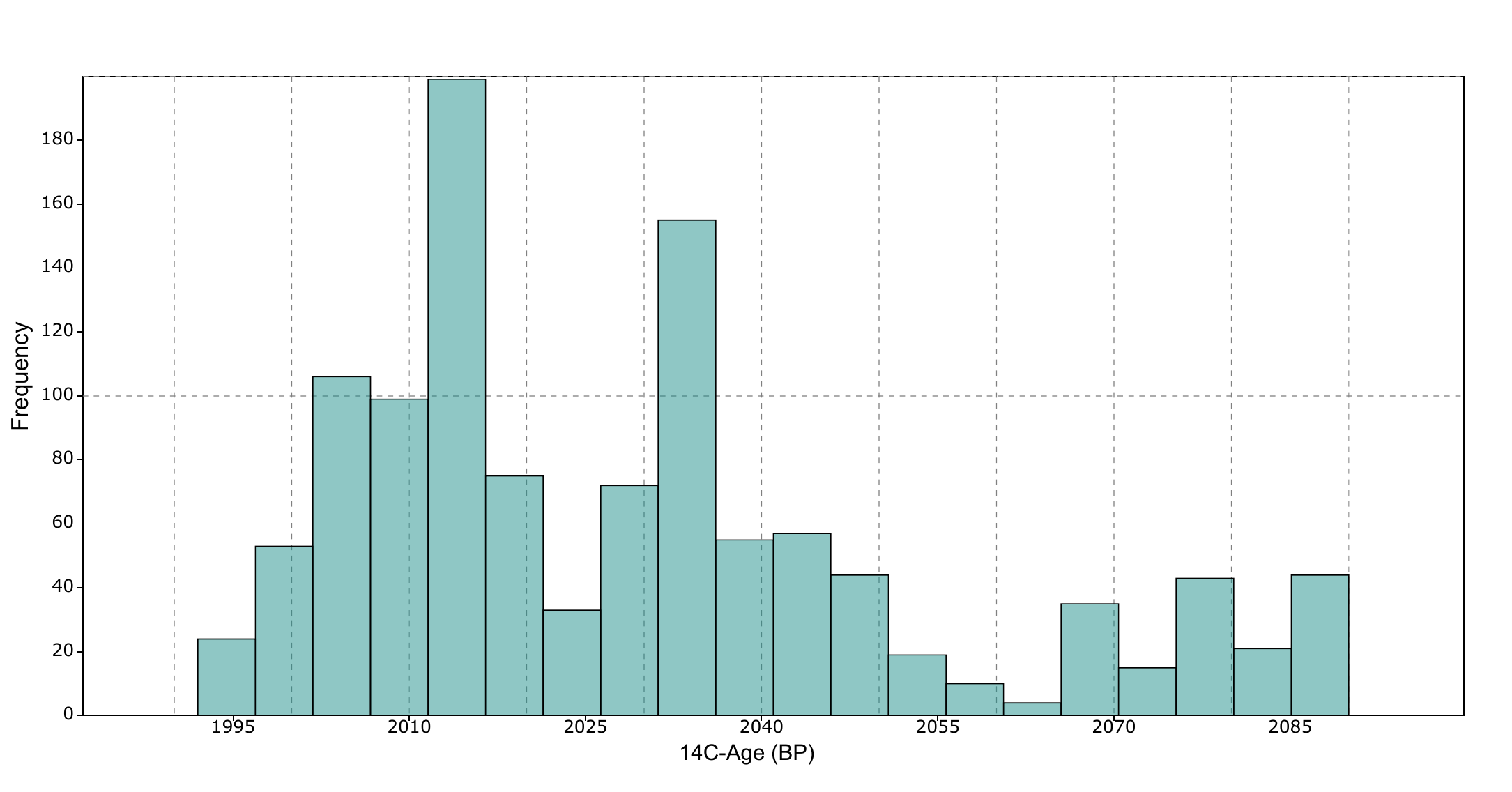}
  \caption{Histogram of the 14C ages found in the reference table 5\_20\_5 for the time point 0 on the floating point timescale; \( n = 1163 \); 22 bins according to Rice's rule (data: worksheet~2 in Appendix~9).}
  \label{fig:fig6}
\end{figure}

Figure 6 presents the histogram of the 1,163 matching 14C age values from the reference table at time point 0. The distribution deviates from normality, exhibiting two distinct peaks---one between approximately 2010 and 2014 BP and another between approximately 2034 and 2036 BP---and spans a range from 1992 to 2090 BP (a total span of 98 radiocarbon years). On average, Test Series 3 yields approximately 1,568 matches per interval (see worksheet 2 in Appendix 9).

Similarly, analysis of the 1,163 corresponding calendar dates reveals an inconsistent distribution, with only 23 unique dates spanning from 95\,BC to AD\,20 (a range of 115 years). Normality of these matches was further evaluated using the Anderson-Darling test (Anderson \& Darling, 1952) via Python’s SciPy package (Virtanen et al., 2020). For the time point 0, a test value of 33.11 was obtained, indicating a statistically significant deviation from normality. Detailed statistics for each interval are provided in Appendix~10 (worksheet 3; see worksheet 4 for normality test per dataset).

In contrast, the distributions of the means and medians within each interval are substantially more concentrated, with the means clustering between approximately 7\,BC and 20\,BC and the medians predominantly falling between 12 and 18\,BC (see Figure~7).

\begin{figure}[htbp]
\centering
  \begin{subfigure}[b]{0.48\textwidth}
    \centering
      \includegraphics[width=\linewidth]{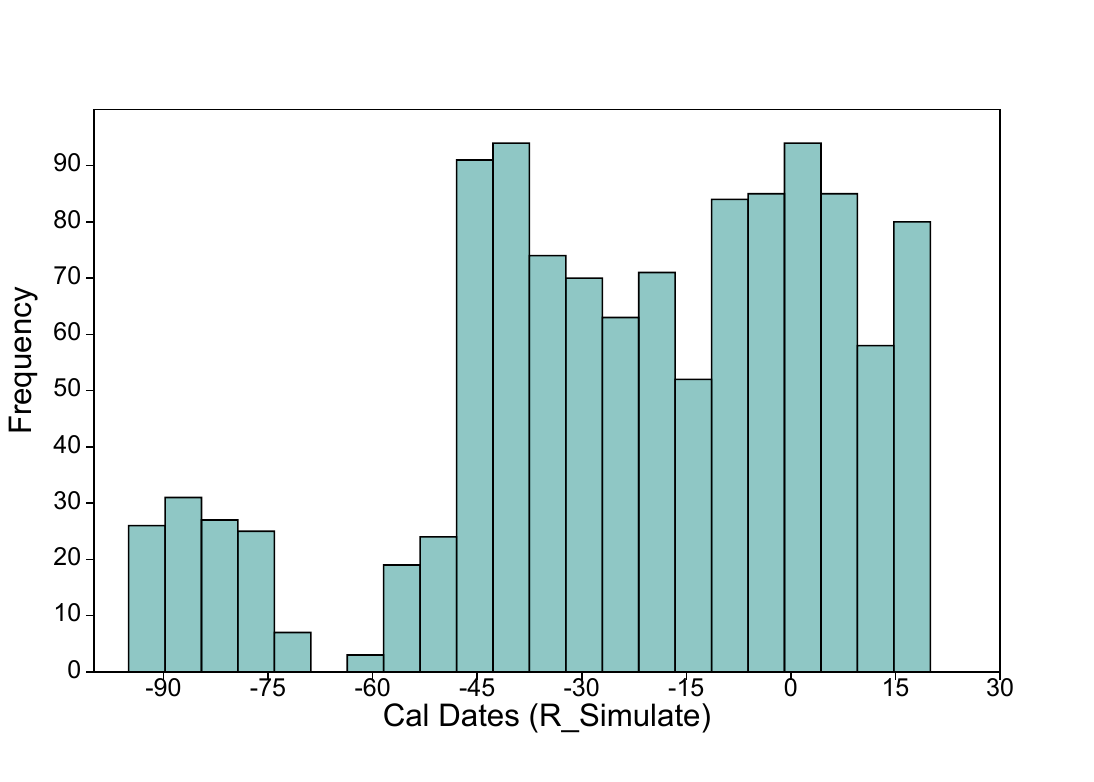}
          \subcaption{Calendar dates}
    \label{fig:7a}
   \end{subfigure}
        \hfill
  \begin{subfigure}[b]{0.48\textwidth}
    \centering
      \includegraphics[width=\linewidth]{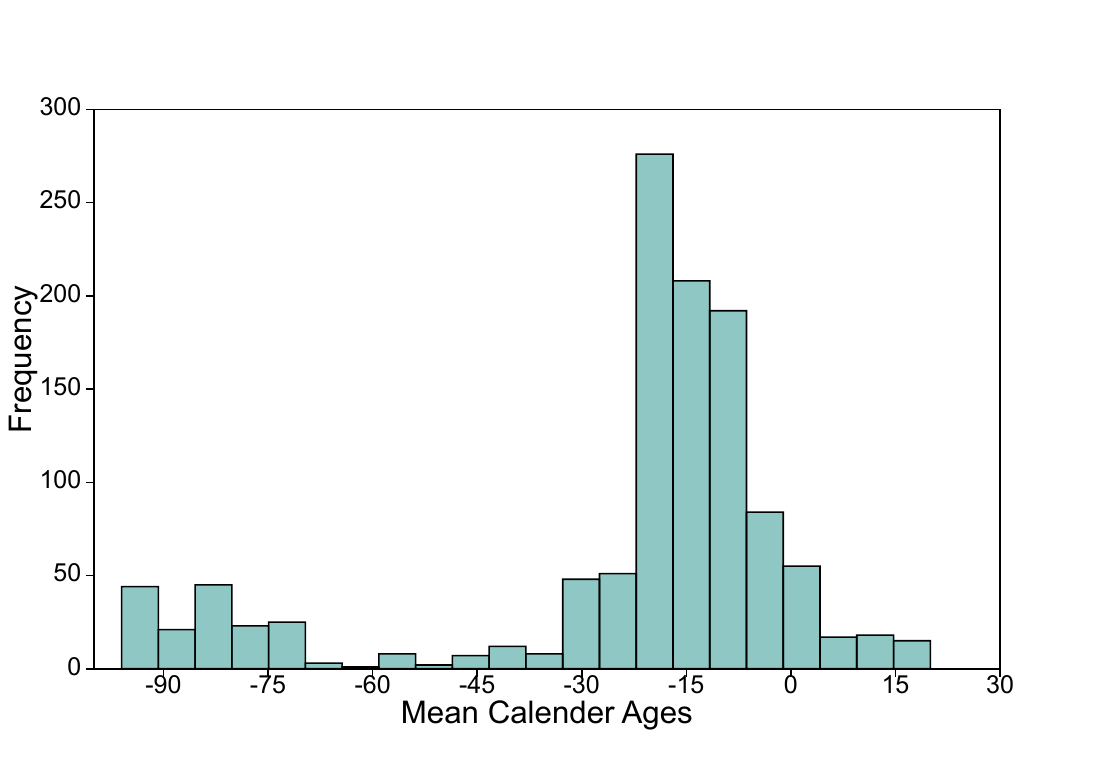}
            \subcaption{Means}
            \label{fig:7b}
    \end{subfigure}
\vspace{1ex}
 \begin{flushleft}
  \begin{subfigure}[b]{0.48\textwidth}
        \raggedright
      \includegraphics[width=\linewidth]{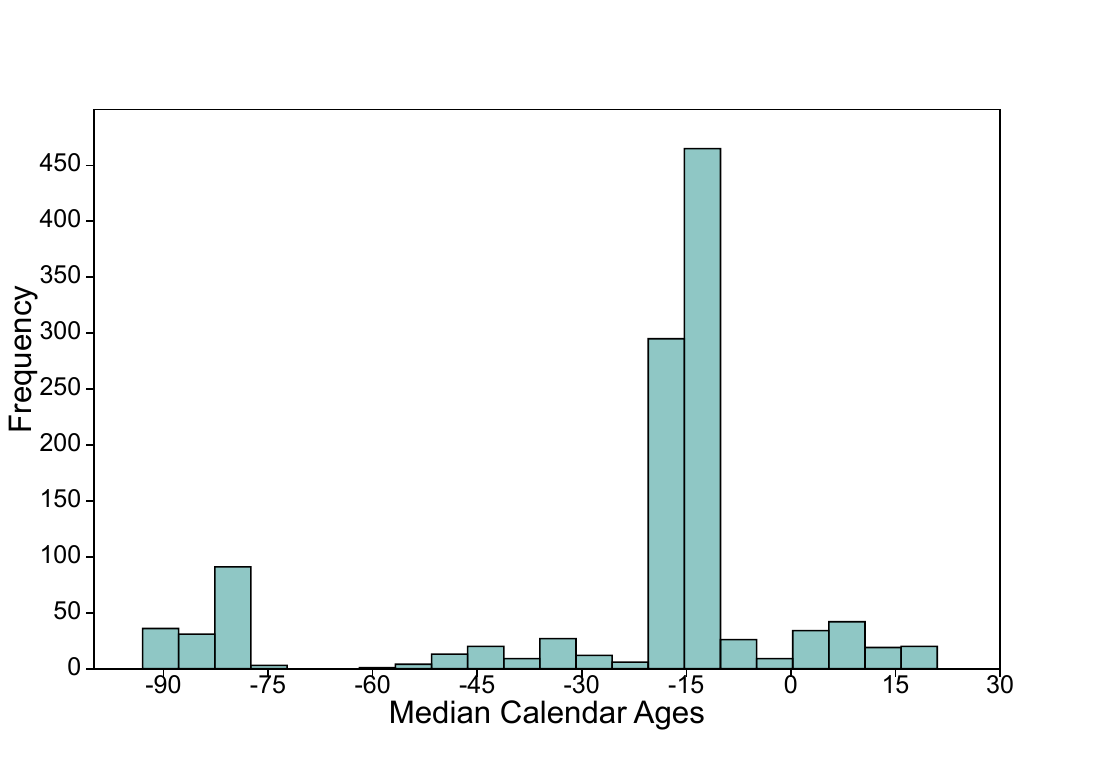}
            \subcaption{Medians}
      \label{fig:7c}
  \end{subfigure}
\end{flushleft}
\caption{Histograms of the three parameters of the fine-dating method for the time interval at time point 0; (a) Calendar dates; (b) Means; (c) Medians. Reference table 5\_20\_5. $n = 1163$; 22 bins according to the Rice rule.}
\label{fig:fig7}
\end{figure}

\subsubsection*{The individual test measurements}
At the level of individual test datasets, as exemplified by the measurements with DataID 2 (test dataset 2) of the Test Series 3 (highlighted in worksheet 2 of Appendix~9), the number of matching values is markedly reduced. In this instance, 12 matching values were identified within reference table 5\_20\_5, corresponding to five unique calendar dates. On average, each test dataset yields approximately 16 records---encompassing the 14C ages and their associated calendar dates, means, and medians---with about half of these calendar dates being unique.

Figure~8 illustrates the histogram of calendar dates for test dataset 2 at time point 0. The dates range from 85\,BC to 20\,AD, with a median of 15\,AD and a mean of approximately 5\,BC, closely approximating the target value of 0.

\begin{figure}[htbp]
\small
  \centering
  \includegraphics[width=0.6\textwidth]{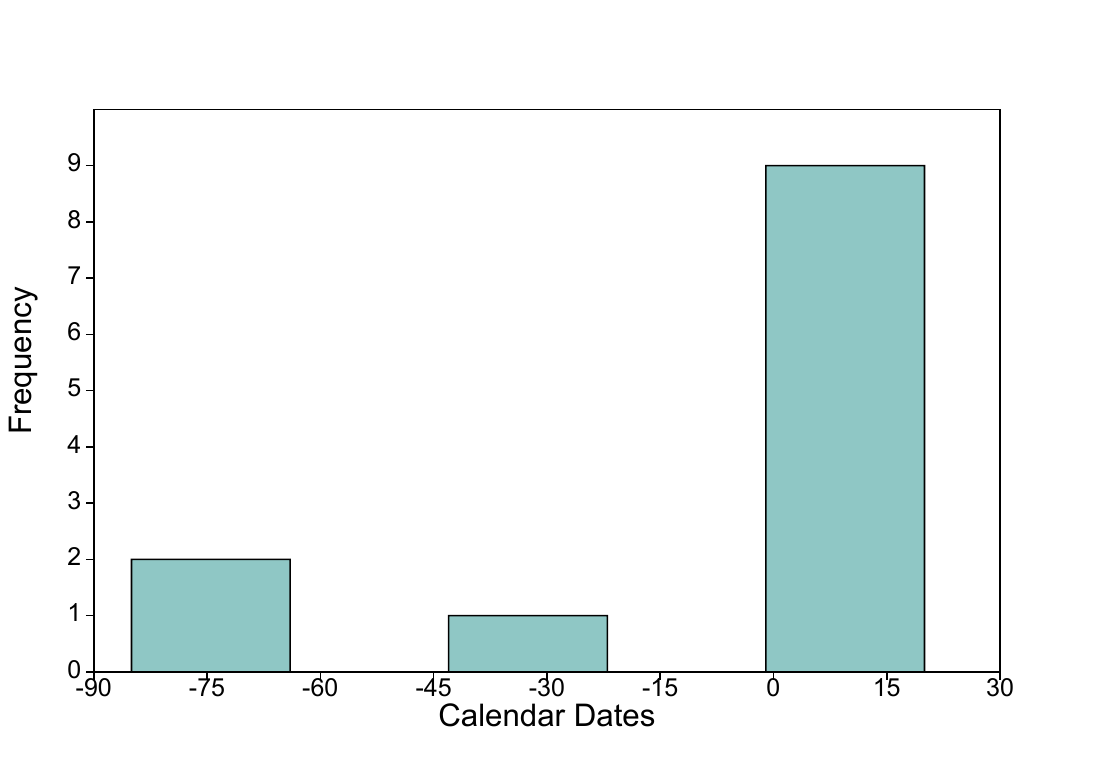}
  \caption{Distribution of calendar date values for test dataset 2 based on reference table 5\_20\_5; time interval: time point 0 on the floating point timescale; \( n = 12 \); 5 bins according to the Rice rule.}
  \label{fig:fig8}
\end{figure}

Concerning the 12 statistical indicators computed by this fine dating method, none of the 12 estimates deviates by more than 24 years from the target; notably, the median of the means and the median of the unique means exhibit deviations of only 1 and 1.5 years, respectively. A comprehensive evaluation across the entire test series further confirms the robustness of these estimates (see worksheet 3 in Appendix~9, with the test dataset 2 highlighted in bold).

\subsubsection*{Overall Performance of the Statistical Indicators --- Delta Comparison}

After confirming the method’s validity and examining its statistical dispersions and interrelationships, we now detail the performance of the twelve statistical indicators across different time periods and reference table variants. Table~4 presents the average deviations (in whole years) of these indicators from the target calendar dates over the entire study period (300\,BC to 0).

\medskip

\begin{table}[htbp]
\centering
\resizebox{\textwidth}{!}{
  \begin{tabular}{l c c c c c c c c c c c c}
  \toprule
\makecell{\textbf{Reference} \\ \textbf{Tab.}} & 
\makecell{\textbf{Avg. $\Delta$} \\ \textbf{CalDate}\\ \textbf{Median}} &
\makecell{\textbf{Avg. $\Delta$} \\ \textbf{CalDate}\\ \textbf{Mean}} &
\makecell{\textbf{Avg. $\Delta$} \\ \textbf{Median}\\ \textbf{Median}} &
\makecell{\textbf{Avg. $\Delta$} \\ \textbf{Median}\\ \textbf{Mean}} &
\makecell{\textbf{Avg. $\Delta$} \\ \textbf{Mean}\\ \textbf{Median}} &
\makecell{\textbf{Avg. $\Delta$} \\ \textbf{Mean}\\ \textbf{Mean}} &
\makecell{\textbf{Avg. $\Delta$} \\ \textbf{unique} \\ \textbf{CalDate}\\ \textbf{Median}} &
\makecell{\textbf{Avg. $\Delta$} \\ \textbf{unique} \\ \textbf{CalDate}\\ \textbf{Mean}} &
\makecell{\textbf{Avg. $\Delta$} \\ \textbf{unique} \\ \textbf{Median}\\ \textbf{Median}} &
\makecell{\textbf{Avg. $\Delta$} \\ \textbf{unique} \\ \textbf{Median}\\ \textbf{Mean}} &
\makecell{\textbf{Avg. $\Delta$} \\ \textbf{unique} \\ \textbf{Mean}\\ \textbf{Median}} &
\makecell{\textbf{Avg. $\Delta$} \\ \textbf{unique} \\ \textbf{Mean}\\ \textbf{Mean}} \\
\midrule
20\_5  & 2 & 2 & -9 & -11 & -12 & -14 & 3 & 2 & -12 & -13 & -15 & -16 \\
50\_5  & 2 & 2 & -11 & -14 & -12 & -15 & 2 & 1 & -16 & -17 & -18 & -19 \\
50\_20 & 2 & 2 & -12 & -13 & -15 & -16 & 2 & 2 & -13 & -15 & -16 & -17 \\
80\_5  & 1 & 1 & -10 & -12 & -12 & -14 & 1 & 1 & -15 & -16 & -17 & -18 \\
100\_0 & 2 & 1 & -11 & -14 & -12 & -14 & 1 & 1 & -16 & -18 & -18 & -19 \\
100\_5 & 2 & 1 & -11 & -13 & -12 & -15 & 1 & 1 & -16 & -18 & -18 & -19 \\
\bottomrule
\end{tabular}
}
\caption{Overview of the average deviations (in years) of the statistical indicators by reference table variant (for full details, see Appendix~11).}
\label{tab:tab5}
\end{table}

\medskip

It is evident that the indicators derived directly from the calendar dates---namely, the median and mean, along with their unique variants---yield the smallest deviations. However, caution is warranted: although deviations as low as one year are observed in some cases, the magnitude and direction of the deviations vary with time. Over the entire study period, these deviations tend to cancel each other out, a phenomenon inherent to the deliberate focus on a 300-year span. A broader temporal scope would likely obscure these fine-scale dynamics and amplify local fluctuations along the calibration curve.

A detailed evaluation of indicator quality is provided in Appendix~11, where performance is categorised as “excellent” (deviation $\leq$10 years), “high-quality” (>10 to $\leq$25 years), “satisfactory” (>25 to $\leq$35 years), or “improvable” (>35 years). For example, based on test measurements using reference table 5\_20\_5, 25.07\% of cases were classified as excellent, 31.12\% as high-quality, 14.53\% as satisfactory, and 29.28\% as improvable. In practical terms, 56.19\% of the fine-dating estimates deviate by no more than 25 years and 70.72\% by no more than 35 years from the target date.

The results further reveal marked differences in performance across specific time periods. Between 300 and 280\,BC, the calculations on the mean and median (and their unique variants) yield high-quality results in over 80\% of cases, whereas estimates based on calendar dates perform less well. 
This discrepancy is likely due to the limited range of the reference table, which extends only until 300~BC, thereby excluding values from earlier periods. As a result, more recent values after 300~BC are given disproportionate weight. From approximately 275\,BC onward, the CalDate estimates are predominantly high-quality. In contrast, the period between 215 and 205\,BC is generally poorly fine-dated, while from 200\,BC onward, the calendar date indicators again exhibit good to very good approximations, despite other indicators performing less favourably. Notably, between 135 and 75\,BC nearly all computational variants are rated as high-quality or excellent, whereas the period from 70 to 50\,BC yields predominantly inferior results. From 45\,BC onward, most fine-dating estimates are high-quality to excellent, with only a slight deterioration from 5\,BC---presumably due to edge-of-scale effects and the absence of compensatory values.

An alternative representation of these results is provided in Figure~9. The higher the line in the chart, the greater the proportion of individual test datasets for which the corresponding parameter group deviated by less than 25 years. Panel (a) shows the proportion of test datasets when using reference table 5\_20\_5, while panel (b) presents the corresponding performance curves for reference table 5\_100\_5. Notably, two problematic regions---around 60\,BC and 210\,BC---are clearly discernible, with substantial differences observed between the individual parameter groups.

\begin{figure}[htbp]
\centering
  \begin{subfigure}[b]{1\textwidth}
    \centering
  \includegraphics[width=0.8\textwidth]{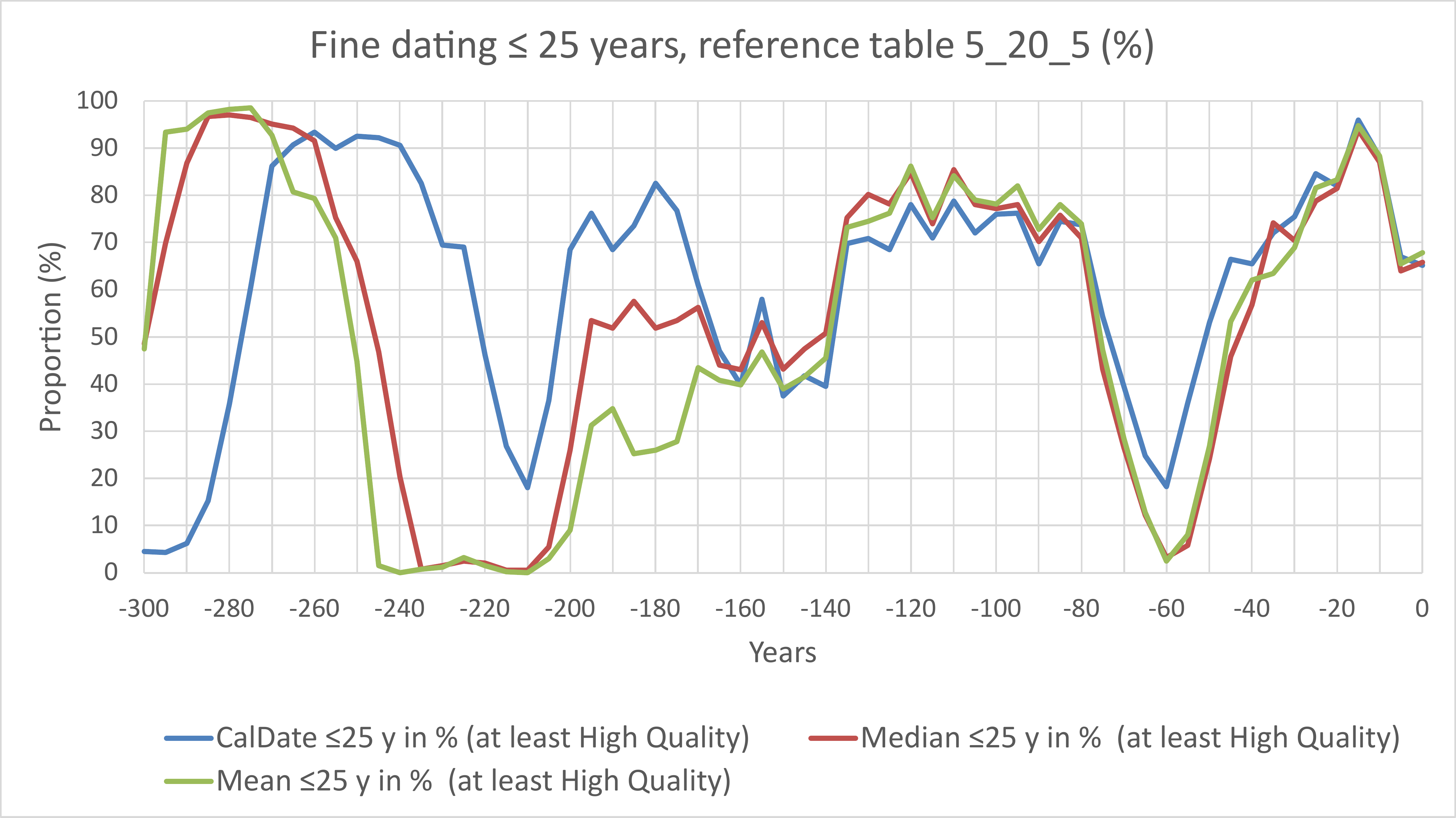}
          \subcaption{Reference table 5\_20\_5}
    \label{fig:9a}
   \end{subfigure}
 \\[0.6cm]
  \begin{subfigure}[b]{1\textwidth}
    \centering
  \includegraphics[width=0.8\textwidth]{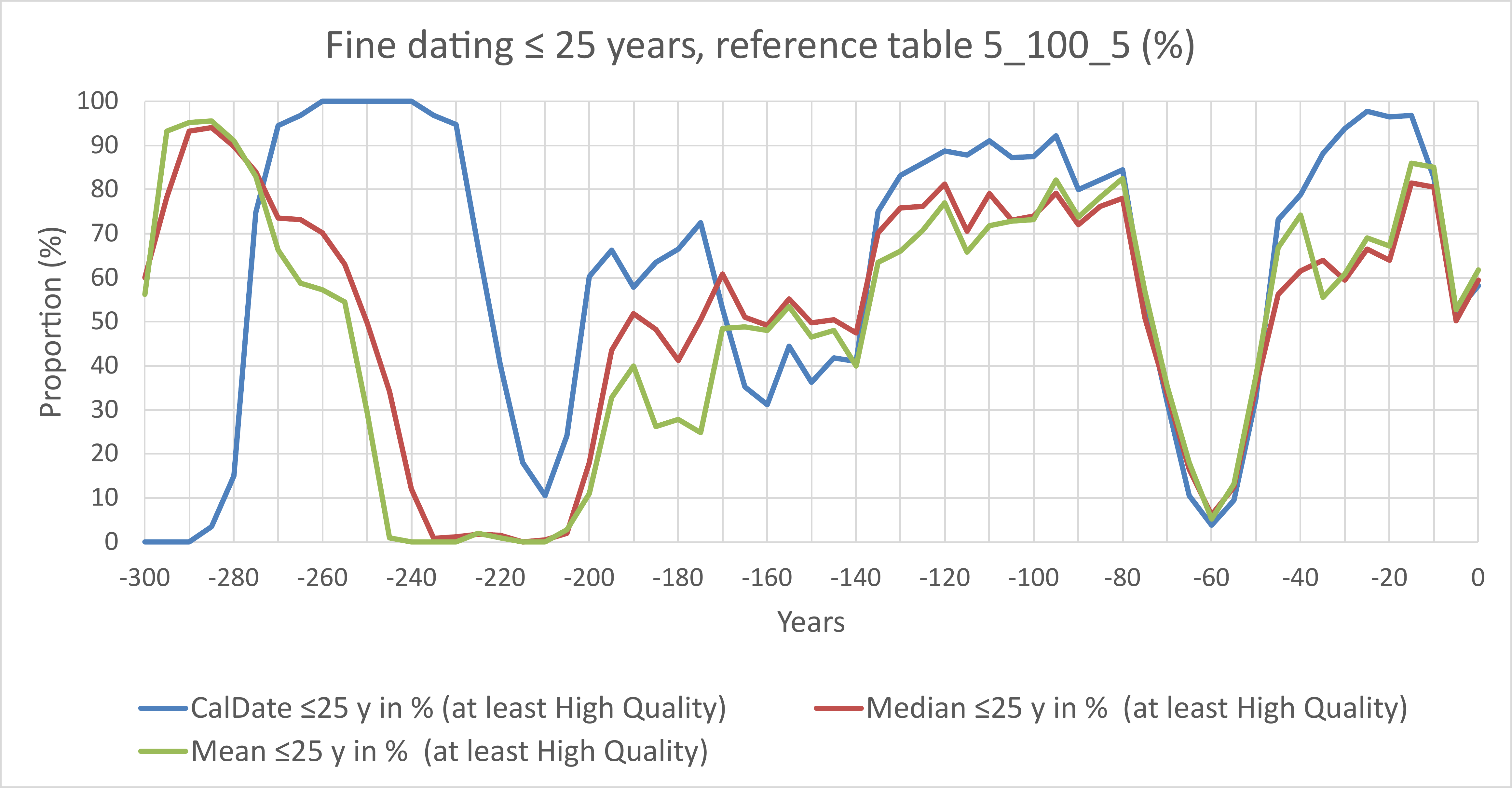}            \subcaption{Reference table 5\_100\_5}
            \label{fig:9b}
    \end{subfigure}
\caption{Performance of the three aggregated parameter groups, based on CalDate, Mean, and Median, for the test dataset (see Appendix 11); criterion: maximum deviation of 25 years. (a) Reference table 5\_20\_5; (b) Reference table 5\_100\_5.}
\label{fig:fig9}
\end{figure}

A comparison of the performance curves reveals two key findings. In certain segments of the study period, minor differences can be observed—with table 5\_100\_5 generally outperforming in most cases. However, the overall trends of both diagrams are almost identical: periods of high and low performance occur at the same time intervals respectively, regardless of the reference table employed, suggesting that the underlying cause lies in the configuration of the calibration curve. As illustrated in Figure~10, regions characterised by a small upward spike adjacent to a minor plateau---specifically around 60\,BC or 210\,BC---exhibit pronounced data imbalances and increased uncertainty. It should be noted, that in areas where the calibration curve is ascending (for example, at 270\,BC), the fine-dating values tend to be younger (since the dispersion in these zones more frequently results in overlapping intervals in the younger segment of the calibration curve).

\begin{figure}[htbp]
  \centering
  \includegraphics[width=0.8\textwidth]{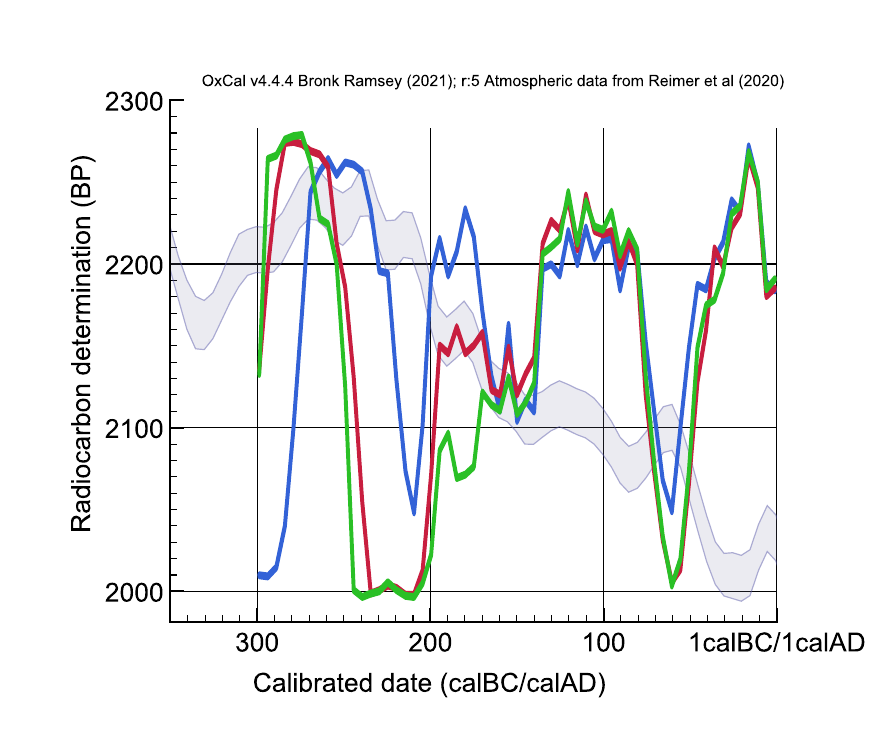}
  \caption{Performance curves of the 5\_20\_5 reference table relative to the IntCal20 calibration curve for the corresponding time interval; criterion used: maximum delta of 25 years.}
  \label{fig:fig10}
\end{figure}

Extending the acceptable deviation to 35 years, the success rate exceeds 80\% across wide segments of the study period. For example, reference table 5\_100\_0 records a 100\% success rate for CalDate estimates between approximately 225 and 270\,BC (see Figure~11). Nevertheless, the problematic areas persist, underscoring the limitations imposed by an unfavourably shaped calibration curve.

\begin{figure}[htbp]
  \centering
  \includegraphics[width=0.8\textwidth]{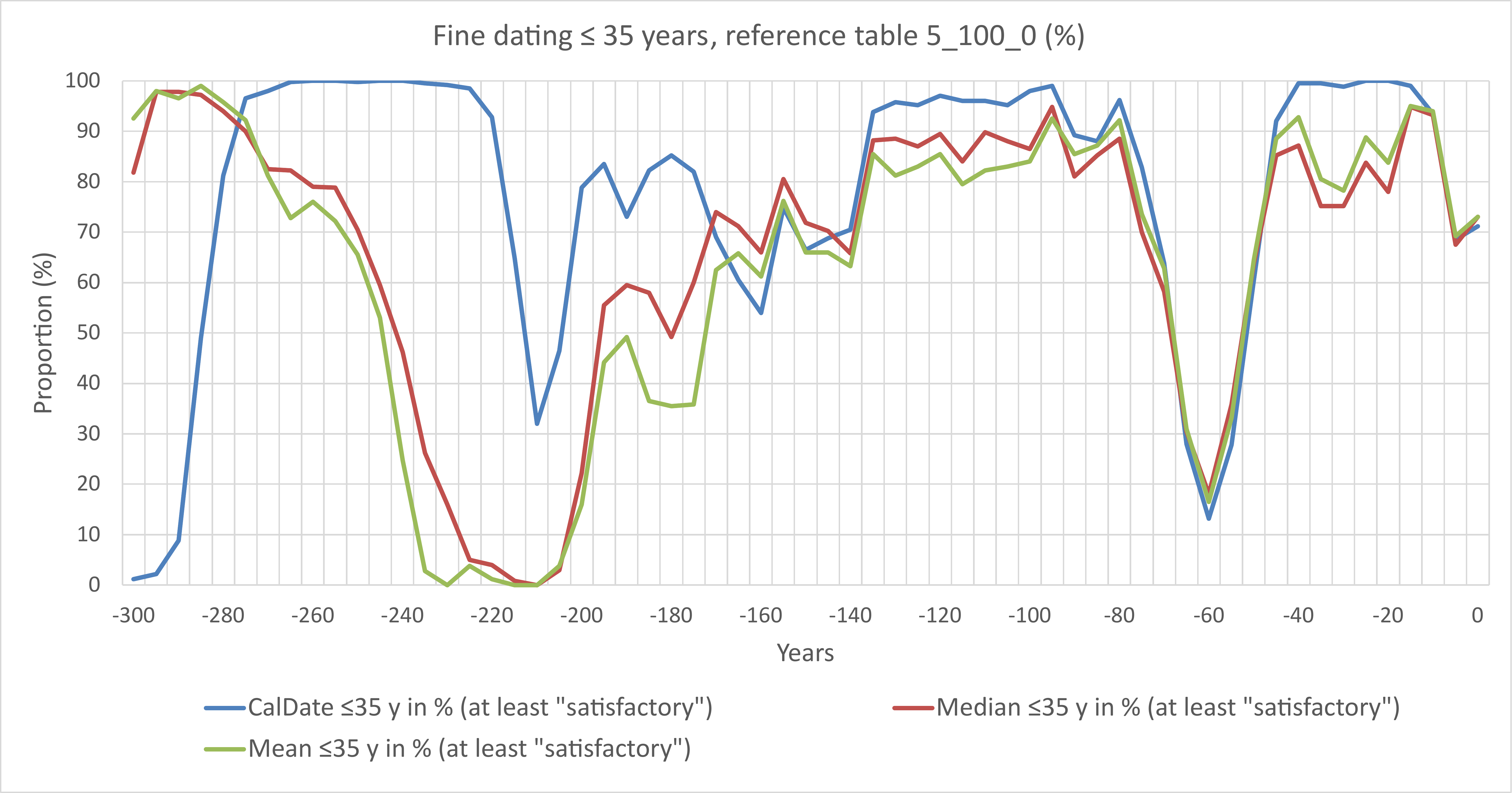}
  \caption{Performance of the three aggregated parameter groups, based on CalDate, Mean, and Median, for the test dataset (see Appendix 11); criteria: maximum deviation of 35 years; reference table 5\_100\_0.}
  \label{fig:fig11}
\end{figure}

Another important aspect of this method relates to the time span covered by the reference table. Test measurements demonstrated significantly improved approximations to the target date when the reference table was restricted to a shorter time span. As shown in Figure~12, the CalDate indicators achieved a 100\% success rate for approximations within a maximum deviation of 25 years over the period from 110\,BC to 80\,BC; in this case, the reference table itself was limited to that interval. This observation is of particular importance, as it highlights the critical impact that the temporal scope of the reference table has on the performance of the fine-dating method. In comparison to calculations performed with the identical data but using a reference table spanning from 1\,BC to 200\,BC (see Appendix~2), a marked improvement is evident.

\begin{figure}[htbp]
  \centering
  \includegraphics[width=0.8\textwidth]{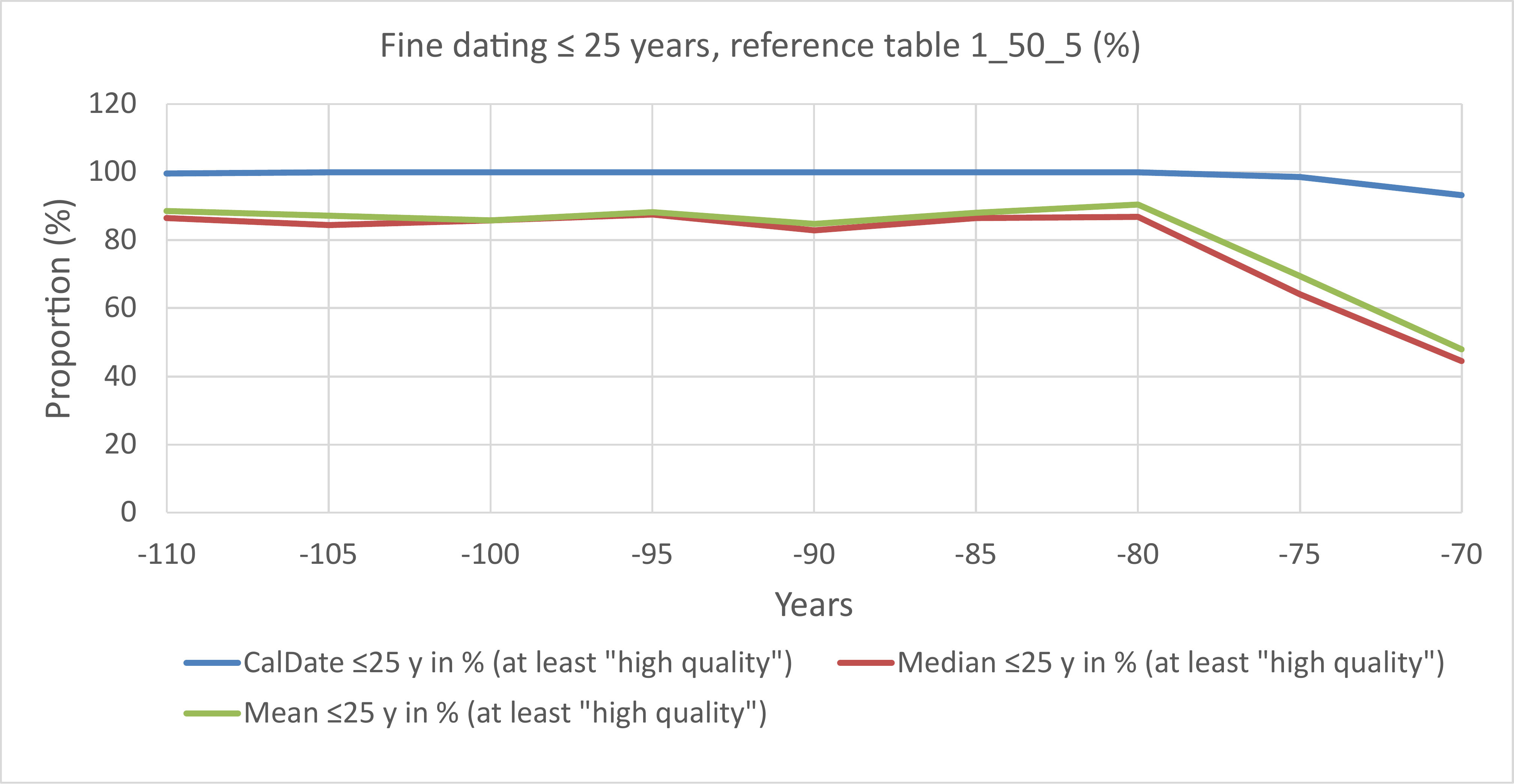}
  \caption{Performance of the three aggregated parameter groups, based on CalDate, Mean, and Median, for the test dataset (see Appendix 11) based on a limited interval of the reference table 1\_50\_5 between 110 and 70~BC; criterion used: maximum delta of 25 years.}
  \label{fig:fig12}
\end{figure}

\noindent
Apart from the method validation, four key findings emerge from this section:
\begin{enumerate}[leftmargin=2em]
    \item The precision of the fine-dating method is highly dependent on the specific time interval and the shape of the calibration curve.
    \item This dependency is consistent across all reference table variants.
    \item There is no clear linear correlation between the data volume of the reference tables and their performance, although minor differences exist.
    \item Significant improvement in performance can be achieved when a chronological framework is known a priori, allowing the reference table’s time span to be restricted and thereby mitigating distortions caused by remote statistical outliers.
\end{enumerate}

\medskip

\subsection{Evaluation of the Fine-Dating Results}

In everyday research practice, one rarely has prior knowledge of the true target calendar date; consequently, examining the deltas offers only a theoretical validation of the method. In practice, when samples fall within problematic regions of the calibration curve, the computed parameters can occasionally yield markedly divergent values. Without a known dating range as a benchmark, such outliers may lead to estimates that are either systematically too young or too old, thereby “contaminating” regions that would otherwise produce reliable approximations.

This issue is exemplified by the comparison between the performance diagram in Figure~11 and the histogram in Figure~13. The histogram, which displays the frequency of computed medians derived from the mean values of the test series based on reference table 5\_100\_0, shows that the x-axis represents the output values rather than the actual underlying calendar dates of the tests.

\begin{figure}[htbp]
  \centering
  \includegraphics[width=0.8\textwidth]{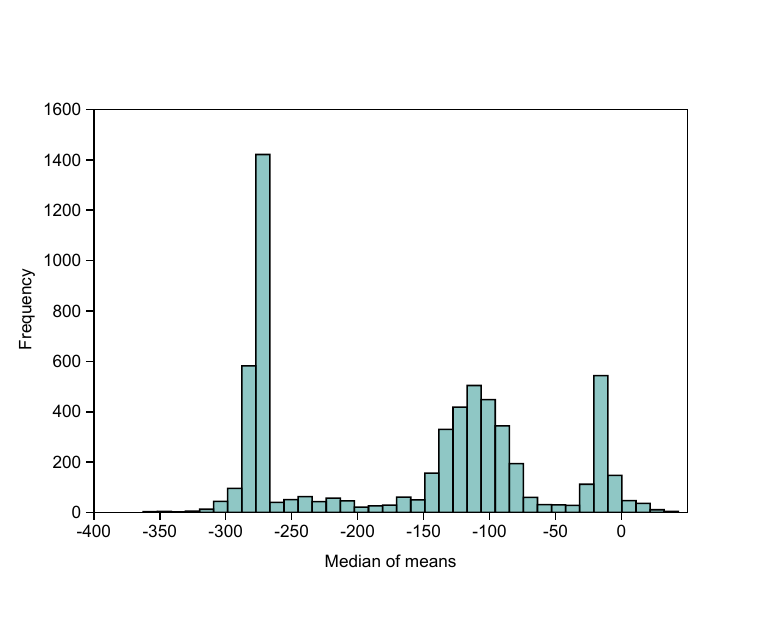}
  \caption{Histogram of the “Mean Median” indicator from the test series (\( n = 6100 \)); reference table 5\_100\_0; 38 bins according to the Rice rule.}
  \label{fig:fig13}
\end{figure}

In this particularly negative example (see Appendix~12 for a full overview of all twelve fine-dating parameters), 1,421 values fall within the bin corresponding to approximately 277--266\,BC, whereas only 40 values are recorded in the adjacent bin (approximately 266--256\,BC). This stark disparity indicates that if a fine-dating process yields a median within the 277--266\,BC range, it is highly likely that the actual target calendar date lies considerably further away; indeed, an examination of the underlying simulation data reveals a dispersion reaching as early as 300\,BC and as late as 180\,BC.

The test series generated for comparative purposes (as described in the Methods section) can also be utilised to visually assess the deviations of individual test parameters. For example, plotting the “CalDate Median” values against the corresponding original calendar dates (see Figure~14) allows one to identify critical regions through their pronounced deviations from the ideal 1:1 line. Although this scatter plot clearly illustrates the shape of the calibration curve (cf. Figures~2 and~10), a drawback of this representation is that the frequency distribution of the many measurements per time interval is not readily discernible; in such cases, one should refer to the histograms presented above.

\begin{figure}[htbp]
  \centering
  \includegraphics[width=0.8\textwidth]{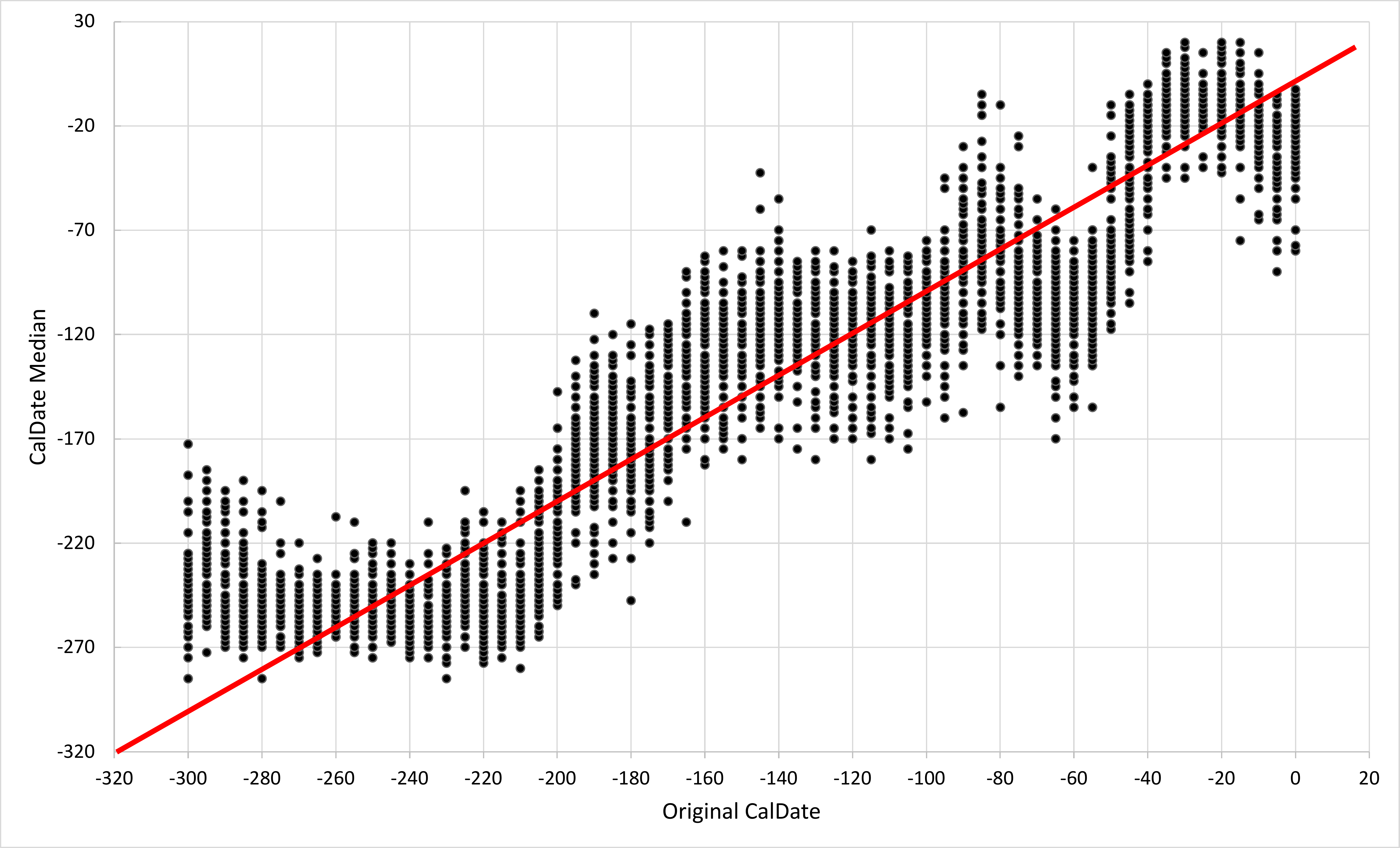}
  \caption{Scatter plot of all measurements of the “CalDate Median” indicator based on a reference table with 50 measurements at 5-year intervals and a SD of 5 years (5\_50\_5), plotted against the original calendar dates.}
  \label{fig:fig14}
\end{figure}

Moreover, the delta comparisons can be generated from simulated reference datasets in routine research practice to provide concrete overviews of potentially dated intervals. The utility of employing a broad spectrum of statistical indicators is further demonstrated by the “Average Deviation Analysis” presented in Appendix~13. In this analysis, colour-coded cells break down the average deviations (deltas) by time interval and indicator, clearly revealing that no single “optimal” indicator exists (and no single reference table as well); rather, the performance of each indicator varies substantially with the time period. Such detailed delta comparisons enable researchers to identify the most suitable indicators for specific chronological phases in fine-dating practice.

Finally, it is important to note that a fundamental understanding of the relevant chronological framework remains essential, as the time values in these analyses refer to the original calendar dates. To facilitate a preliminary assessment of the fine-dating results---even in the absence of original calendar dates---a lookup table was developed (see Appendix~14), in which the years correspond directly to the values in the individual columns.

\section{Discussion}

The method presented here harnesses simulated data from OxCal to substantially expand the effective sample size for radiocarbon fine-dating. By directly linking uncalibrated radiocarbon ages with calendar dates through repeated simulations, our approach enables the calculation of robust central tendency measures (e.g. medians and means) directly from the dispersion of simulated calendar dates, and further aggregates these measures in a bootstrapping-like manner. While we hold established methods such as Bayesian modelling in high regard for their rigour and versatility, our fine-dating method is designed to serve as a valuable complement within the existing methodological canon.

Reproducibility and transparency are cornerstones of our approach. The generation of simulated datasets using OxCal’s robust \texttt{R\_Simulate} function---which has proven its reliability in multiple studies---is supported by a suite of Python scripts that automate the extraction, matching, and statistical evaluation of radiocarbon ages. Furthermore, the independent validation provided by dendrochronologically dated tree rings underpins the method’s credibility; these real-world data confirm that the simulated outputs closely mirror empirical measurements, ensuring that the method is both practical and scientifically robust. The wide range of statistical indicators employed also permits a nuanced evaluation of performance across different segments of the calibration curve, thus facilitating targeted improvements in regions with known challenges.

Despite its strengths, the method is not without limitations. It is inherently sensitive to the shape of the calibration curve; regions characterised by plateaus or abrupt fluctuations---such as the problematic zones around 60\,BC and 210\,BC---tend to produce heterogeneous distributions of simulated values. In these areas, the dispersion is markedly higher, and the resulting estimates may systematically deviate from the true calendar dates. Although the use of multiple statistical indicators and an increased sample size can partially mitigate these effects, further refinements are necessary.

Looking ahead, several avenues for improvement and further research are evident. Refinement of the reference tables---optimising both the number of measurements per interval and the temporal buffer zones---is essential to enhance performance in problematic regions of the calibration curve. Future comparative studies combining Bayesian and simulation-based approaches may yield hybrid methods that harness the strengths of both paradigms, ultimately providing a more reliable and nuanced chronological framework.

\section{Conclusion}

In summary, our study has introduced a novel method that utilises simulated data from OxCal’s \texttt{R\_Simulate} function to achieve high-resolution fine-dating of radiocarbon measurements.

The method demonstrates that, through repeated simulations and a bootstrapping-like aggregation of statistical indicators, it is possible to derive calendar dates with a precision that is comparable to, and in some contexts even exceeds, that obtained via conventional Bayesian methods. Importantly, the approach is conceived as a complement to the existing radiocarbon dating toolbox---offering an accessible, transparent, and reproducible alternative that requires only minimal additional computational effort.

The method’s strength is underscored by its empirical validation using both extensive simulated test datasets and independently dendrochronologically dated tree rings, which confirm that the simulated outputs closely mirror actual measurements. This accessibility and robustness make the approach particularly attractive for researchers seeking high-resolution chronological reconstructions without the need for extensive prior information.

Looking forward, further refinements are expected to enhance the method’s performance even in challenging segments of the calibration curve. These methodological considerations not only address current limitations but also pave the way for targeted future investigations.

Ultimately, our fine-dating method represents a significant step forward in absolute dating, providing a transparent, efficient, and empirically sound tool that can be systematically improved and broadly applied across diverse scientific disciplines beyond archaeology, including geochronology, palaeoclimatology, and forensic science.

\section{Acknowledgements}

We gratefully acknowledge the contributions of several colleagues who provided invaluable support during the course of this study. We thank Prof. Dr. Sabine Hornung for her insightful discussions, meticulous proofreading, and, most importantly, for inspiring my engagement with the radiocarbon method and Bayesian modeling. We are equally indebted to Prof. Dr. Bernd Ulmann for his thorough review of the manuscript, with particular attention to the mathematical notations. Our sincere thanks also extend to Mr. Erik Reischl for his proofreading, as well as for carefully reviewing and commenting on the Python code. We appreciate the many stimulating conversations, editorial suggestions, and continuous motivation provided by Ms. Birte Voltmer. Finally, we acknowledge Dr. Bernd Oehme for his detailed proofreading of the text.

\section{References}

\begin{description}[leftmargin=2em]
      \item[Aitken, M.J. (1990)] Science‐based dating in archaeology. London and New York: Longman Arch. Ser. \href{https://doi.org/10.4324/9781315836645}{doi:10.4324/9781315836645}
    \item[Anderson, T.W. and Darling, D.A. (1952)] Asymptotic theory of certain ‘goodness‐of‐fit’ criteria based on stochastic processes. \textit{The Annals of Mathematical Statistics}, 23(2), pp.193--212. \href{https://doi.org/10.1214/aoms/1177729437}{doi:10.1214/aoms/1177729437}
    \item[Barral, P. and Fichtl, S. (eds.) (2012)] Regards sur la chronologie de la fin de l'âge du Fer (IIIe--Ier Jahrhundert vor Chr.) en Gaule non méditerranéenne. Actes de la table ronde tenue à Bibracte, 15--17 octobre 2007. Glux-en-Glenne: Coll. Bibracte 22.
    \item[Bayliss, A. (2009)] Rolling out revolution: using radiocarbon dating in archaeology. \textit{Radiocarbon}, 51(1), pp.123--147. \href{https://doi.org/10.1017/S0033822200033750}{doi:10.1017/S0033822200033750}
    \item[Bayliss, A. (2015)] Quality in Bayesian chronological models in archaeology. \textit{World Archaeology}, 47(4), pp.677--700. \href{https://doi.org/10.1080/00438243.2015.1067640}{doi:10.1080/00438243.2015.1067640}
    \item[Berger, R. and Suess, H.E. (eds.) (1979)] Radiocarbon dating. Proceedings of the Ninth International Conference, Los Angeles and La Jolla, 1976. Berkeley, Los Angeles, London: UC Press Voices Revived.
    \item[Bronk Ramsey, C. (1995)] Radiocarbon calibration and analysis of stratigraphy: the OxCal program. \textit{Radiocarbon}, 37(2), pp.425--430. \href{https://doi.org/10.1017/S0033822200030903}{doi:10.1017/S0033822200030903}
    \item[Bronk Ramsey, C. (1998)] Probability and dating. \textit{Radiocarbon}, 40, pp.461--474.
    \item[Bronk Ramsey, C. (2001)] Development of the radiocarbon calibration program. \textit{Radiocarbon}, 43(2A), pp.355--363. \href{https://doi.org/10.1017/S0033822200038212}{doi:10.1017/S0033822200038212}
    \item[Bronk Ramsey, C. (2009a)] Bayesian analysis of radiocarbon dates. \textit{Radiocarbon}, 51(1), pp.337--360. \href{https://doi.org/10.1017/S0033822200033865}{doi:10.1017/S0033822200033865}
    \item[Bronk Ramsey, C. (2009b)] Dealing with outliers and offsets in radiocarbon dating. \textit{Radiocarbon}, 51(3), pp.1023--1045. \href{https://doi.org/10.1017/S0033822200034093}{doi:10.1017/S0033822200034093}
    \item[D'Agostino, R. (1971)] An omnibus test of normality for moderate and large sample sizes. \textit{Biometrika}, 58(2), pp.341--348.
    \item[D'Agostino, R. and Pearson, E.S. (1973)] Tests for departure from normality. Empirical results for the distributions of $b_2$ and $\sqrt{b_1}$. \textit{Biometrika}, 60(3), pp.613--622.
    \item[Demir, S. (2022)] Comparison of normality tests in terms of sample sizes under different skewness and kurtosis coefficients. \textit{International Journal of Assessment Tools in Education}, 9(2), pp.397--409. \href{https://doi.org/10.21449/ijate.1101295}{doi:10.21449/ijate.1101295}
    \item[Fahrni, S.M. et al. (2020)] Single-year German oak and Californian Bristlecone Pine 14C data at the beginning of the Hallstatt Plateau from 856\,BC to 626\,BC. \textit{Radiocarbon}, 62(4), pp.919--937. \href{https://doi.org/10.1017/RDC.2020.16}{doi:10.1017/RDC.2020.16}
    \item[Fürst, S. and Hornung, S. (2024)] Radiokarbondatierungen. In: Hornung, S. (ed.) Das spätrepublikanische Militärlager von Hermeskeil (Lkr. Trier-Saarburg). Studien zur frühen römischen Präsenz im Treverergebiet. Dossiers Arch. (Centre Nat. Recherche Arch.) 21 = Saarbrücker Beitr. Altkde. 91. Bertrange, pp.651--670.
    \item[Gazoni, E. and Clark, C. (2023)] openpyxl --- A Python library to read/write Excel 2010 xlsx/xlsm files. Version 3.1.2. Available at: \url{https://openpyxl.readthedocs.io/} (Accessed: 20 November 2024).
    \item[Hamilton, W.D. and Krus, A.M. (2018)] The myths and realities of Bayesian chronological modeling revealed. \textit{American Antiquity}, 83(2), pp.187--203. \href{https://doi.org/10.1017/aaq.2017.57}{doi:10.1017/aaq.2017.57}
    \item[Hamilton, D., Haselgrove, C. and Gosden, C. (2015)] The impact of Bayesian chronologies on the British Iron Age. \textit{World Archaeology}, 47, pp.642--660. \href{https://doi.org/10.1080/00438243.2015.1053976}{doi:10.1080/00438243.2015.1053976}
    \item[Harris, C.R. et al. (2020)] Array programming with NumPy. \textit{Nature}, 585(7825), pp.357--362. \href{https://doi.org/10.1038/s41586-020-2649-2}{doi:10.1038/s41586-020-2649-2}
    \item[Higham, T. and Petchey, F. (2000)] Radiocarbon dating in archaeology: methods and applications. In: Creagh, D.C. and Bradley, D.A. (eds.) (2000) Radiation in art and archeometry. Amsterdam, Lausanne, New York: n.p., pp.255--284.
    \item[Holland-Lulewicz, J. and Ritchison, B.T. (2021)] How many dates do I need? \textit{Advances in Archaeological Practice}, 9(4), pp.272--287. \href{https://doi.org/10.1017/aap.2021.10}{doi:10.1017/aap.2021.10}
    \item[Libby, W.F. (1955)] Radiocarbon dating. Chicago: University of Chicago Press.
    \item[McKinney, W. (2010)] Data structures for statistical computing in Python. In: Proceedings of the 9th Python in Science Conference, pp.51--56.
    \item[Michczyński, A. (2007)] Is it possible to find a good point estimate of a calibrated radiocarbon date? \textit{Radiocarbon}, 49(2), pp.393--401. \href{https://doi.org/10.1017/S0033822200042326}{doi:10.1017/S0033822200042326}
    \item[Python Software Foundation (2024)] The Python Standard Library: os module documentation. Available at: \url{https://docs.python.org/3/library/os.html} (Accessed: 20 November 2024).
    \item[Reimer, P.J. et al. (2020)] The IntCal20 Northern Hemisphere radiocarbon age calibration curve (0–55 cal kBP). \textit{Radiocarbon}, 62(4), pp.725--757. \href{https://doi.org/10.1017/RDC.2020.46}{doi:10.1017/RDC.2020.46}
    \item[Renfrew, C. (1973)] Before civilisation. The radiocarbon revolution and prehistoric Europe. London: Penguin Books.
    \item[Renfrew, C. and Bahn, P.G. (2016)] Archaeology: theories, methods and practice. London: Thames \& Hudson.
    \item[SciPy (2024)] \texttt{scipy.stats.normaltest} (version 1.14.1). Available at: \url{https://docs.scipy.org/doc/scipy/reference/generated/scipy.stats.normaltest.html} (Accessed: 20 November 2024).
    \item[Shapiro, S.S. and Wilk, M.B. (1965)] An analysis of variance test for normality (complete samples). \textit{Biometrika}, 52(3--4), pp.591--611.
    \item[Stuiver, M., Reimer, P.J. and Braziunas, T.F. (1998)] High-precision radiocarbon age calibration for terrestrial and marine samples. \textit{Radiocarbon}, 40(3), pp.1127--1151. \href{https://doi.org/10.1017/S0033822200019172}{doi:10.1017/S0033822200019172}
    \item[Taylor, R.E. and Bar-Yosef, O. (2014)] Radiocarbon dating: an archaeological perspective. Walnut Creek, CA: Left Coast Press.
    \item[Virtanen, P. et al. (2020)] SciPy 1.0: fundamental algorithms for scientific computing in Python. \textit{Nature Methods}, 17, pp.261--272. \href{https://doi.org/10.1038/s41592-019-0686-2}{doi:10.1038/s41592-019-0686-2}
    \item[Wood, R. (2015)] From revolution to convention: the past, present and future of radiocarbon dating. \textit{Journal of Archaeological Science}, 56, pp.61--72. \href{https://doi.org/10.1016/j.jas.2015.02.019}{doi:10.1016/j.jas.2015.02.019}

\end{description}

\section{Appendices}

\begin{enumerate}[leftmargin=2em]
    \item Functionality of \texttt{R\_Simulate} in OxCal
    \item Reference Tables
    \item Python-Script \texttt{R\_Sim\_to\_R\_Date}
    \item Python-Script \texttt{R\_Date}
    \item Report 75BC based on Refrence Table 5\_20\_5
    \item (a) Report file for Python-Script "ReferenceDataSets\_Parameters"\\
          (b) Reference file for Python-Script "ReferenceDataSets\_Parameters" \\
          (c) Python-Script "ReferenceDataSets\_Parameters"
    \item Python-Script "ReferenceDataSets\_Lookup"
    \item Dendrochronologically dated measurements
    \item TestSeries3: Simulated Test Data
    \item TestSeries3: Normality Tests
    \item Delta-Categories
    \item Histograms Fine Dating 5\_100\_0
    \item Average Deviation Analysis
    \item LookupTable based on reference table 5\_50\_5
\end{enumerate}

\end{document}